\documentclass[]{ACTA} % don't know what the [] does

\usepackage{newtxtext,newtxmath} % I like this more
\usepackage{microtype}
\usepackage[english]{babel}
\usepackage[utf8]{inputenc}
\usepackage{bm}
\usepackage{amsmath}
\usepackage{float}
\usepackage{hyperref}
\usepackage{cite}
\usepackage{tabularray}
\usepackage{enumitem}
\usepackage[switch,columnwise]{lineno}
\modulolinenumbers[5]
\usepackage{gensymb}
\usepackage{fancyhdr}
 \hypersetup{
    colorlinks,
    linkcolor={black!50!},
    citecolor={black},
    urlcolor={black}}

\begin{document}

% \IACpaperyear{23}
% \IACpapernumber{C1.5.2}
% \IACconference{74}
% \IACcopyright{74th}{2023}{Chiara Pozzi}{Mauro Pontani}
% \IAClocation{Baku, Azerbaijan}

% \title{Trajectory Optimization, Guidance and Control of Low-Thrust Orbit Transfers from the Lunar Gateway to Low Lunar Orbit}

\title{Optimization, Guidance, and Control of Low-Thrust Transfers from the Lunar Gateway to Low Lunar Orbit}

\IACauthor{Chiara Pozzi}{Department of Aerospace Engineering, Khalifa University of Science and Technology, Abu Dhabi, P.O. Box 127788, United Arab Emirates; 100064456@ku.ac.ae\\
Faculty of Civil and Industrial Engineering, Sapienza Università di Roma, Rome, Italy;  
pozzi.1791751@studenti.uniroma1.it}
\IACauthor{Mauro Pontani}{Department of Astronautical, Electrical, and Energy Engineering, Sapienza Università di Roma, Rome, Italy; mauro.pontani@uniroma1.it}

\IACauthor{Alessandro Beolchi}{Department of Aerospace Engineering, Khalifa University of Science and Technology, Abu Dhabi, P.O. Box 127788, United Arab Emirates; 100064448@ku.ac.ae \\
Faculty of Civil and Industrial Engineering, Sapienza Università di Roma, Rome, Italy; beolchi.1802082@studenti.uniroma1.it}

\IACauthor{Elena Fantino}{Department of Aerospace Engineering, Khalifa University of Science and Technology, Abu Dhabi, P.O. Box 127788, United Arab Emirates; elena.fantino@ku.ac.ae}

\abstract{The Lunar Gateway will represent a primary space system useful for the Artemis program, Earth-Moon transportation, and deep space exploration. It is expected to serve as a staging location and logistic outpost on the way to the lunar surface. This study focuses on low-thrust transfer dynamics, from the Near-Rectilinear Halo Orbit traveled by Gateway to a specified Low-altitude Lunar Orbit (LLO). More specifically, this research addresses two closely-related problems: (i) determination of the minimum-time low-thrust trajectory and (ii) design, implementation, and testing of a guidance and control architecture, for a space vehicle that travels from Gateway to LLO. Orbit dynamics is described in terms of modified equinoctial elements, with the inclusion of all the relevant perturbations, in the context of a high-fidelity multibody ephemeris model. The minimum-time trajectory from Gateway to a specified lunar orbit is detected through an indirect heuristic approach, which uses the analytical conditions arising in optimal control theory in conjunction with a heuristic technique. However, future missions will pursue a growing level of autonomy, and this circumstance implies the mandatory design and implementation of an efficient feedback guidance scheme, capable of compensating for nonnominal flight conditions. This research proposes nonlinear orbit control as a viable and effective option for autonomous explicit guidance of low-thrust transfers from Gateway to LLO. This approach allows defining a feedback law that enjoys quasi-global stability properties without requiring any offline reference trajectory. The overall spacecraft dynamics is modeled and simulated, including attitude control and actuation. The latter is demanded to an array of reaction wheels, arranged in a pyramidal configuration. Guidance, attitude control, and actuation are implemented in an iterative scheme. Monte Carlo simulations demonstrate that the guidance and control architecture at hand is effective in nonnominal flight conditions, i.e. with random starting point from Gateway as well as in case of temporary unavailability of the propulsion system. The numerical results also point out that only a modest propellant penalty is associated with the use of feedback guidance and control in comparison to the minimum-time optimal trajectory. \\

}

\maketitle \thispagestyle{fancy}

%\linenumbers

\section{{Introduction}}
The Lunar Gateway is a primary component of NASA Artemis program as it will play a crucial role as a habitation and logistic outpost along the way to the lunar surface. The space station will be assembled and operational in the vicinity of the Moon and shall be used as a central location for aggregation of supplies and resources for human missions in cislunar space and beyond. Gateway will travel a $L_2$ Near-Rectilinear Halo Orbit (NRHO) of the southern family with a 9:2 lunar synodic resonant period. The chosen orbit offers a reasonably low perilune radius, which proves beneficial for surface access \cite{zimovan2017characteristics}. Specifically, the resulting orbit exhibits an average period of 6.6 days, with a perilune radius ranging from 3196 to 3557 km and an apolune of approximately 68000 km above the south pole. This configuration allows for an extended communication period between Gateway and potential facilities located at the lunar south pole. Perfectly periodic in the Circular Restricted 3-Body Problem (CR3BP) model, the NRHO exists as a quasi-periodic orbit in a higher-fidelity model. These types of orbits are set apart from the rest of the halo orbits by their favorable stability characteristics \cite{DavisBoudad}. Gateway NRHO trajectory is collected in an SPK-type kernel compatible with the SPICE ephemeris system created at the NASA Jet Propulsion Laboratory (JPL) \cite{lee2018sample}. \\
\indent In recent years, low-thrust electric propulsion is gaining increasing relevance and already found application in a variety of mission scenarios \cite{rayman,RATHSMAN}. Low-thrust transfers require continuous thrusting, leading to a continuous control history that defines the thrust direction and magnitude at each instant.  One approach to addressing this is incorporating optimization methods into the process. The benefits that optimization methods provide are numerous, but they also bring with them new challenges. Spacecraft motion is governed by a system of nonlinear differential equations and, incorporating low-thrust forces into this system, adds a new layer of challenges in constructing desirable trajectories. This results in many more design variables and less intuitive problems overall.  Low-thrust trajectory optimization problems are solved using direct, indirect, and heuristic approaches, sometimes combined in hybrid form \cite{pontminfue}. Achieving an optimal solution is a time-demanding endeavor and can only be completed offline. However, actual mission scenarios often entail real-time demands for which low-thrust propulsion can serve as an effective solution.\\
\indent To address real-time low-thrust orbit guidance, researchers have explored various methods in the literature. Some have utilized the variational equations \cite{kluever,gurfilgf}, and others used a combination of Cartesian coordinates and orbit elements in the dynamical framework of formation flying \cite{Schaub}. Another approach is the neighboring optimal guidance \cite{PONTANIcecchetti}, which stands out for its ability to minimize the additional propellant required for correcting nonnominal flight conditions. However, the effectiveness of a neighboring optimal guidance technique relies on having a reference path in addition to all associated state and costate variables. Identifying such an optimal reference trajectory in certain dynamical scenarios, such as orbit maintenance in the presence of perturbations or unmodeled dynamics, can be demanding or even infeasible. This underscores the importance of developing an orbit control strategy designed to compensate for orbit perturbations and temporary engine failures during transfer orbits.\\ 
\indent In the context of cislunar space trajectories, numerous transfer design applications have exploited the inherent manifold structures in the framework of the restricted three-body problem. These structures facilitate departure and arrival through nearly ballistic trajectories \cite{HOWELL2006367,Topputo2005}. In cases where manifold structures are not leveraged, alternative strategies, encompassing low-thrust or impulsive maneuvers, have been explored to address orbit transfers. Extensive research has been conducted on NRHOs, including transfer missions between NRHOs and various orbit types such as Low Earth Orbits (LEO), Distant Retrograde Orbits (DRO), and Low Lunar Orbits (LLO) \cite{may2020enabling,TROFIMOV2020260,Tseulo}. Both robotic and manned orbit transfers between Gateway and LLOs are essential for the future Artemis Base Camps on the lunar surface. Therefore, the design of transfer trajectories from NRHOs to LLOs becomes pivotal. Numerous research in the literature have treated direct transfers between NRHOs and LLOs involving several high-thrust maneuvers \cite{LuLin,thangavelu}. Notable studies have devised direct trajectories from halo orbits to LLOs employing Hohmann transfers as a preliminary estimate for identifying alternative paths \cite{cao}. However, to the best of the author's knowledge, no existing research focuses on low-thrust transfers between the NRHO traveled by Gateway and a LLO. This gap in the existing literature stimulates the development of an orbit control strategy designed to effectively compensate for orbit perturbations and overcome challenges related to the utilization of low-thrust engines. Attitude control is also crucial for the design of a more realistic mission scenario. The full 6-degree-of-freedom motion was explored by Knutson and Howell, considering a spacecraft moving along Lyapunov and Halo orbits \cite{Knutson}. Additionally, existing literature focused on guidance, control, and navigation for a 6-degree-of-freedom spacecraft in the context of cislunar rendezvous and docking. \cite{ sanchez2020chance, colagrossi2021guidance, muralidharan2023rendezvous }.\\
\indent In this work, low-thrust transfer trajectories from Gateway to low-altitude polar operational orbits are designed and analyzed. The choice of a LLO as the arrival orbit is driven by strategic considerations. This orbital choice facilitates transfers with the lunar surface, providing improved access to desired landing sites. Notably, LLOs with high inclinations are especially suitable for missions to the lunar poles, offering convenient access to sites crucial for future Artemis base camps. In this work, the previously mentioned gaps are addressed. This study has the following objectives: (i) develop an indirect optimization scheme based on optimal control theory and compatible with a multiple-body dynamical framework, to find the low-thrust optimal transfer, (ii) design an effective autonomous guidance technique based on Lyapunov stability theory, (iii) incorporate attitude control and actuation in the dynamical modeling, for the purpose of designing and testing a comprehensive guidance, control, and actuation architecture. The latter is intended to represent a new methodology for autonomous guidance and control and does not require any offline (nominal) trajectory. However, the minimum-time transfer (objective (i)) is being identified, for the purpose of comparing the performance of the overall architecture with the theoretical limit.\\
\indent This work is organized as follows. In Section \ref{sect2}, the reference frames, the spacecraft state representation, and the governing equations for both orbit and attitude dynamics are introduced. Then, the minimum-time transfer path is investigated in Section \ref{sect3}. Finally, in Section \ref{sect4}, the derivation of nonlinear feedback control laws for both orbit and attitude, along with their application to the problem at hand, is discussed. This section also emphasizes the crucial issue of the autonomy of space missions.

\section{{Spacecraft dynamics}}\label{sect2}
To explore strategies for designing trajectories for vehicles moving in a multi-body dynamical scenario, the very first step consists in understanding and modeling the environment in which these spacecraft will travel. Various reference frames must be introduced. Moreover, the equations of motion, both for orbit and attitude dynamics need to be discussed.
\subsection{Reference Frames}
The Earth-Centered Inertial frame (ECI) has its origin at the Earth's center and is associated with vectrix
\begin{equation}\label{ECI}
    \underline{\underline{E}} = [{\hat{c}_1}^{E} \: \: {\hat{c}_2}^{E} \: \: {\hat{c}_3}^{E}]
\end{equation}
\noindent where unit vectors ${\hat{c}_1}^{E}$ and ${\hat{c}_2}^{E}$ lie on the Earth's mean equatorial plane. In particular, ${\hat{c}_1}^{E}$ is the vernal axis, while ${\hat{c}_3}^{E}$ points toward the Earth rotation axis.\\
\indent The Moon-Centered Inertial frame (MCI) is defined in relation to the ECI frame, it has the origin in the Moon's center and is associated with vectrix
\begin{equation}\label{MCI}
    \underline{\underline{M}} = [{\hat{c}_1}^{M} \: \: {\hat{c}_2}^{M} \: \: {\hat{c}_3}^{M}]
\end{equation}
\noindent where unit vectors ${\hat{c}_1}^{M}$ and ${\hat{c}_2}^{M}$ define the Moon's mean equatorial plane, which is assumed to coincide with the ecliptic plane. In particular, ${\hat{c}_1}^{M}$ is chosen to be parallel to ${\hat{c}_1}^{E}$ (i.e. the vernal axis), while ${\hat{c}_3}^{M}$ points toward the Moon rotation axis that is perpendicular to the ecliptic plane. The MCI frame is defined in relation to the ECI frame through the ecliptic obliquity angle ($\epsilon = 23.4$ degree)
\begin{equation}\label{MCIMAT}
    \underline{\underline{M}}^T = \textbf{R}_1(\epsilon)\underline{\underline{E}}^T
\end{equation}
\noindent where $\textbf{R}_j(\xi)$ denotes an elementary counterclockwise rotation about axis $j$ by a generic angle $\xi$.\\
\indent The synodic reference frame is a non-inertial frame with origin in the center of mass of the Earth-Moon system. It is associated with vectrix
\begin{equation}\label{SYNfr}
    \underline{\underline{R}}_{SYN} = [\hat{i} \: \: \hat{j} \: \: \hat{k}]
\end{equation}
\noindent where $\hat{i}$ points toward the Moon along the line joining the two massive bodies, $\hat{k}$ points toward the Moon orbit angular momentum, and $\hat{j}$ is such that the triad $\left({\hat{i}}\: \: {\hat{j}} \: \: {\hat{k}}\right)$ is a right-handed sequence of unit vectors. \\
\indent The Local-Vertical Local-Horizontal frame (LVLH) rotates together with the space vehicle and is defined in relation to a main attracting body. The frame is associated with vectrix
\begin{equation}\label{LVLH}
    \underline{\underline{R}}_{LVLH} = [\hat{r} \: \: \hat{\theta} \: \: \hat{h}]
\end{equation}
\noindent where $\hat{r}$ is aligned with the spacecraft position vector, taken from the center of the main attracting body, $\hat{h}$ points toward the spacecraft orbit angular momentum, and $\hat{\theta}$ is chosen such that ($\hat{r}$, $\hat{\theta}$, $\hat{h}$) is a right-handed sequence of unit vectors. The LVLH frame has its origin in the center of mass of the spacecraft itself. If $\Omega$, $i$ and $\theta_t$ denote respectively the instantaneous right ascension of the ascending node (RAAN), inclination and argument of latitude, then 
\begin{equation}\label{LVLHmat}
    \underline{\underline{R}}_{LVLH}^T = \textbf{R}_3(\theta_t)\textbf{R}_1(i)\textbf{R}_3(\Omega)\underline{\underline{M}}^T
\end{equation}
\noindent where $\underline{\underline{M}}$ is the MCI frame.

\subsection{Orbit Dynamics}
Orbit dynamics is described in terms of Modified Equinoctial Elements (MEE), an alternative orbital element set that avoids singularities in the event of circular or equatorial orbits. MEE are defined in terms of the Classical Orbit Elements (COE), i.e. semimajor axis $a$, eccentricity $e$, inclination $i$, RAAN $\Omega$, argument of periapsis $\omega$, and true anomaly $\theta_*$,
\begin{equation}\begin{split}\label{COE2NEE}
p &= a \, (1-e^2) \, \, \, \, \, \, \, \, \, \, \, \, \, \, \, l = e \, \cos{(\Omega+\omega)}\\
m &= e \, \sin{(\Omega+\omega)} \, \, \, \, \, \, n = \tan{\frac{i}{2}}\cos{\Omega} \\
s &= \tan{\frac{i}{2}}\sin{\Omega} \, \, \,\, \, \, \, \, \, \, \, \, q = \Omega + \omega + \theta_{*}.
\end{split}\end{equation}
\noindent This set is nonsingular except for equatorial retrograde orbits ($i = \pi)$, which is a case that seldom occurs. \\ 
\indent Denoting ${\bm z} = \begin{bmatrix}
        x_1 & x_2 & x_3 & x_4 & x_5
    \end{bmatrix}^T \equiv \begin{bmatrix}
        p & l & m & n & s
    \end{bmatrix}^T$ and $x_6 \equiv q$, the governing equations for the MEE are 
    \begin{equation}\label{zeq}
\begin{split}
     \dot{\bm z} &= {\bm G}\left({\bm z}, x_6\right){\bm a} \\
     \dot{x_6} &= {\sqrt \frac{\mu}{x_1^3}} \, \eta^2 + {\sqrt \frac{x_1}{\mu}} \, \frac{x_3 \, \sin{x_6} - x_5 \, \cos{x_6}}{\eta} \, a_{h}
\end{split}
\end{equation}
\noindent where ${\bm a} = \begin{bmatrix}
        a_r & a_\theta & a_h
    \end{bmatrix}^T$ is the non-Keplerian acceleration in the LVLH frame, which includes both the thrust and the perturbing acceleration, whereas $\eta = 1 + x_2 \cos{x_6} + x_3 \sin{x_6}$. Moreover, ${\bm G}\left({\bm z}, x_6\right)$ is a $5 \times 3$ matrix defined as
    
    \begingroup
\thinmuskip=0mu
\medmuskip=0mu
\thickmuskip=0mu
\begin{equation}\label{Gmat}
     \resizebox{.88\hsize}{!}{${\bm G}\left({\bm z}, x_6\right)=\sqrt{\cfrac{x_1}{\mu}}{\begin{bmatrix}
       0 & \cfrac{2x_1}{\eta} & 0 \\
       s_{x_6} & \cfrac{(\eta+1)c_{x_6}+x_2}{\eta} & -\cfrac{x_4 s_{x_6}-x_5 c_{x_6}}{\eta}x_3 \\
       -c_{x_6}
&  \cfrac{(\eta+1)s_{x_6}+x_3}{\eta} &   \cfrac{x_4 s_{x_6}-x_5 c_{x_6}}{\eta}x_2 \\
0 & 0 & \cfrac{1+x_4^2+x_5^2}{2\eta} c_{x_6}\\
0 & 0 & \cfrac{1+x_4^2+x_5^2}{2\eta} s_{x_6}
\end{bmatrix}}$}.
\end{equation}
\endgroup

\noindent 

The boundary conditions of the problem to be introduced are assumed to involve only the first five MEE, collected in ${\bm z}$.

\indent The dynamics of the space vehicle is described by the state vector $\bm x$ defined as
\begin{equation}\label{stateveccomplete}
    {\bm x} = \begin{bmatrix}
        {\bm z}^T & x_6 & x_7
    \end{bmatrix}^T
\end{equation}
\noindent where the last component is the mass ratio $x_7= \frac{m}{m_0}$, and its governing equation is 
\begin{equation}\label{x7doteq}
    \dot{x}_7 = -\frac{u_T}{c}
\end{equation}
\noindent where $c$ is the constant effective exhaust velocity of the propulsion system, while $u_T$ is the ratio between the thrust magnitude and the initial mass
\begin{equation}\label{uttt}
    u_T = \frac{T}{m_0} \, \, \, \, \, \, \text{with } \, \, \, \, \, 0 \leq u_T \leq u_T^{(max)}.
\end{equation}
\noindent The instantaneous thrust acceleration can be expressed as 
\begin{equation}
    {\bm a}_T = \frac{{\bm u}_T}{x_7}
\end{equation}
\noindent with magnitude constrained in the interval $0 \leq a_T \leq a_T^{(max)}$.\\
The state equations (\ref{zeq}) and (\ref{x7doteq}) can be written in compact form as 
\begin{equation}\label{constr1}
        \bm{\dot{x}}=\bm{f}\left(\bm{x}, \, \bm{u}, \, t\right).
\end{equation}

\subsection{Third Body Gravitational Perturbation}
\indent When a spacecraft orbits a primary celestial body, the gravitational influence of additional massive objects, commonly known as third bodies, can be considered as a perturbation acting on the vehicle. The gravitational action of the Earth and the Sun must be taken into account. Denoting the spacecraft with $S$, the dominating body with $1$, the third body with $3$, and neglecting the mass of the space vehicle, the gravitational perturbation exerted on $S$ due to body $3$ is
\begingroup
\thinmuskip=0mu
\medmuskip=0mu
\thickmuskip=0mu
\begin{equation}\label{a3B}
    \underrightarrow{\boldsymbol{a}}_{3B} = \mu_3\,\left\{\frac{\underrightarrow{\boldsymbol{r}}_{13} - \underrightarrow{\boldsymbol{r}}_{1S}}{\left[\left( \underrightarrow{\boldsymbol{r}}_{1S} -\underrightarrow{\boldsymbol{r}}_{13} \right)\cdot\left( \underrightarrow{\boldsymbol{r}}_{1S} -\underrightarrow{\boldsymbol{r}}_{13} \right)\right]^{\frac{3}{2}}} - \frac{\underrightarrow{\boldsymbol{r}}_{13}}{r_{13}^3}\right\}
\end{equation}
\endgroup
\noindent where $\mu_3$ is the gravitational parameter of the third body, $\underrightarrow{\boldsymbol{r}}_{1j}$ with 
 $j = 3$ or $S$ is the position vector of the j-th body relative to the primary, and $r_{1j} = |\underrightarrow{\boldsymbol{r}}_{1j}|$.\\
\indent The instantaneous position vectors of the Earth and the Sun are extracted from the ephemeris, therefore a higher-fidelity model serves as the dynamical basis for the analysis. The JPL's Horizon ephemeris model \cite{acton2018spice} is used in this study. Table \ref{tab1} reports the gravitational parameters of the relevant celestial bodies.

\bgroup
\def\arraystretch{1.2}
\begin{table}[h] 
\centering
\begin{tabular}{cc}
\toprule
\text{Celestial Object} & \text{${\mu} \, \, \left[{\frac{km^3}{s^2}}\right]$} \\
\midrule
Sun&1.3271 $\cdot 10^{11}$\\
Earth&3.9860 $\cdot 10^5$\\
Moon&4.9028 $\cdot 10^3$\\
\bottomrule
\end{tabular}
\vspace*{3mm}
\caption{Gravitational parameters of the relevant celestial objects}
\label{tab1}
\end{table}
\egroup

\subsection{Attitude Dynamics}
Assuming that the spacecraft is a rigid body, a right-hand sequence, termed body frame, is attached to it, 
\begin{equation}\label{bodyframe}
      \underline{\underline{B}} = [{\hat{b}_1} \: \: {\hat{b}_2} \: \: {\hat{b}_3}].
 \end{equation}
The attitude of a space vehicle can be described using quaternions $ \{q_0, \, {\bm{q}}\}$. The same attitude can also be represented by the rotation matrix, which links the inertial frame MCI to the body frame, written in terms of $\{q_0, \, {\bm{q}}\}$, 
\begin{equation}
    \underset{\rm B \leftarrow M}{\rm \textbf{R}} = \left(q_0^2 - \bm{q}^T  \bm{q}\right) \, I_{3x3} \, + 2 \, \bm{q} \, \bm{q}^T - 2 \, q_0 \, \bm{\tilde{q}}
\end{equation}
\noindent where 
\begin{equation}
   \bm{\tilde{q}}  = \begin{bmatrix} 0 & -q_3 & q_2 \\ \\ q_3 & 0 & -q_1\\ \\ -q_2 & q_1 & 0 \end{bmatrix}.
\end{equation}

Let $\bm{\omega}$ be the (3 $\times$ 1)-vector that collects the three components (along the three axes) of the spacecraft angular velocity with respect to the inertial frame, the kinematics equations that govern the time evolution of the quaternions are
\begin{equation}\label{qpunto}
        \dot{q}_0 = -\frac{1}{2} \, \bm{\omega}^T\bm{q} \, \, \, \, \, \, \,  \dot{\bm{q}} = \frac{1}{2} \,\left[q_0 \, \bm{\omega} + \tilde{\bm{q}} \bm{\omega}\right].
\end{equation}
The dynamics equations of a rigid body, namely Euler's equation of motion, are
\begin{equation}\label{eulerdef}
    \bm{J}_c^{(B)}\bm{\dot{\omega}} +\, \bm{\tilde{\omega}}\bm{J}_c^{(B)}\bm{{\omega}} \, = \, \bm{T}_c + \bm{M}_c
\end{equation}
\noindent where the matrix $\bm{J}_c^{(B)}$ is the inertia matrix about the center of mass resolved in the body axes frame (\ref{bodyframe}), $\bm{T}_c$ is the control torque and $\bm{M}_c$ is the resultant of the external torques ($3 \times 1$ vectors with components along the body axes frame). 

\section{{Minimum-time orbit transfer}}\label{sect3}
This section is devoted to investigating the minimum-time transfer path from the initial orbit of Gateway to a final low-altitude lunar orbit. In this context, attitude dynamics is neglected for the purpose of identifying the minimum-time path, and the vehicle is modeled as a point mass. The initial time (i.e., the starting point on the NRHO) is not specified and the objective function $J$ to minimize is 
\begin{equation}\label{costfunc}
    J = K_J\left(t_f -t_0\right)
\end{equation}
\noindent where $K_J$ is a positive constant chosen arbitrarily. \\
\indent Previous research \cite{pontaniMEE} demonstrated that the use of MEE considerably mitigates the hypersensitivity of the numerical solution to the initial values of the adjoint variables.
\subsection{Formulation of the problem}
The space vehicle of interest is governed by the state equations (\ref{constr1}) and is subject to problem-dependent boundary conditions written in vector form as
\begin{equation}\label{constr2}
 \bm{{\Psi}}\left(\bm{x}_0, \, \bm{x}_f,\, t_0, \, t_f\right) = {\bm 0}.
\end{equation}
\noindent In this research, the state vector of the departing Gateway orbit is a function of the initial time $t_0$, which is not fixed. Thus, the initial boundary conditions are defined once the starting time is known
\begin{equation}
    \bm{\Psi}_0=  \begin{bmatrix}
       x_{10}-p(t_0) \\
       x_{20}-l(t_0)\\
       x_{30}-m(t_0)\\
       x_{40}-n(t_0)\\
       x_{50}-s(t_0)\\
       x_{60}-q(t_0) \\
       x_{70}-1
    \end{bmatrix} = \bm{0}.
\end{equation}
\noindent In this study, only three of the six possible classical orbital elements are prescribed for the final orbit, i.e. the semilatus rectum $p_d$, the eccentricity $e_d$, and the inclination $i_d$. For the final boundary condition vector, one gets
\begin{equation}\label{finalbound2}
    \bm{\Psi}_f=  \begin{bmatrix}
       x_{1f}-p_d \\ 
       x_{2f}^2+x_{3f}^2-e_d^2\\ 
       x_{4f}^2+x_{5f}^2-\tan^2{\cfrac{i_d}{2}}
    \end{bmatrix} = \bm{0}.
\end{equation}
The overall boundary condition vector is $\bm{\Psi}=  \left[\bm{\Psi}_0^T \, \, \, \, \bm{\Psi}_f^T\right] = \bm{0}^T$.

\subsection{Necessary conditions for optimality}
To state the necessary conditions for optimality, an auxiliary function $\Phi$ and the Hamiltonian function $H$ are introduced  (see also \cite{hull2013optimal})
\begin{equation}
   \Phi=K_J\left(t_f -t_0\right) + \bm{\nu}^T\bm{\Psi}
\end{equation}
\begin{equation}\label{definitionofH}
    H = \bm{\lambda}^T\bm{f}
\end{equation}
\noindent where $\bm{\nu}$ is the time-independent adjoint vector used as conjugate to $\bm{\Psi}$, whereas $\bm{\lambda}(t)$ is the time-varying adjoint vector, also termed costate vector, associated with the state equations. \\
\indent The necessary conditions for optimality can be derived after lengthy developments, omitted for the sake of brevity \cite{pontanibook},  
    \begin{equation} \label{NCgeneral}
    \begin{split}
        &\bm{\dot{\lambda}} = -\left(\frac{\partial H}{\partial \bm{x}}\right)^T  \, \, \, \, \, \, \, \, \,\bm{u} = {\rm argmin}_{\bm{u}}H\\
        &\bm{\lambda}_0 = -\left(\frac{\partial \Phi}{\partial \bm{x}_0}\right)^T  \, \, \, \, \, \,  \bm{\lambda}_f = \left(\frac{\partial \Phi}{\partial \bm{x}_f}\right)^T  \\
        &H_0 = \frac{\partial \Phi}{\partial t_0} \, \, \, \, \, \, \, \, \, \, \, \, \, \, \, \, \, \, \,  H_f = -\frac{\partial \Phi}{\partial t_f}.
        \end{split}
    \end{equation}

From the Pontryagin Minimum Principle (PMP), the optimal thrust direction that minimizes the Hamiltonian is obtained,
\begin{equation}
  \begin{aligned}\label{anglesalfa}
        \sin{\alpha} &= -\frac{H_r}{\sqrt{H_r^2+H_{\theta}^2}} \, \, \, \, \, \, 
        \cos{\alpha} = -\frac{H_\theta}{\sqrt{H_r^2+H_{\theta}^2}}
  \end{aligned}
\end{equation}
\begin{equation}\label{anglesbeta}
 \sin{\beta} = -\frac{H_h}{\sqrt{H_r^2+H_{\theta}^2+H_h^2}}\\
\end{equation}
\noindent where $\alpha$ is the angle in the ($\hat{r}, \, \hat{\theta}$) plane with $-\pi \leq \alpha \leq \pi$, while $\beta$ is the out-of-plane angle such that $-\frac{\pi}{2} \leq \beta \leq \frac{\pi}{2}$.
\begin{figure}[htp]
\centering
\includegraphics[width=\columnwidth]{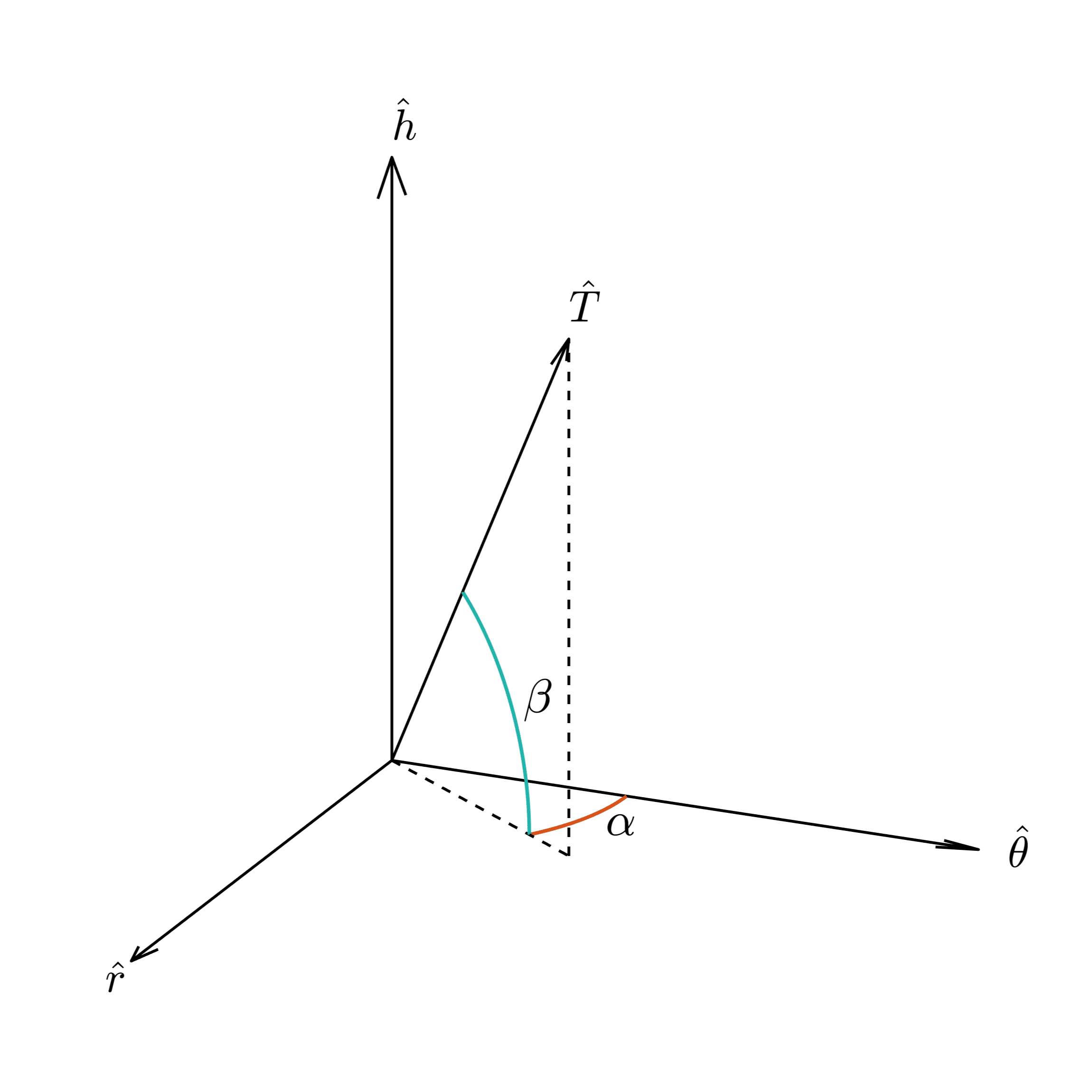}
    \caption{Thrust angles}
    \label{anglesthrust}
\end{figure}
\noindent Through the use of PMP and the necessary conditions in Eq. (\ref{NCgeneral}), it is possible to demonstrate that the thrust magnitude must be set to its maximum \cite{pontanibook} (cf. Eq. (\ref{uttt})). Hence, the equation governing the evolution of the mass ratio over time can be replaced by a linear decreasing function of time 
\begin{equation}\label{x7anal}
    x_7 \left( t \right) = 1 - \frac{u_T^{(max)}}{c} \, (t-t_0).
\end{equation}
This avoids using the mass ratio and thrust-to-initial mass ratio (i.e. $u_T$) as state and control components, respectively, in the optimization process. Consequently, the Hamiltonian function can be rewritten  (see \cite{pontanibook})
\begin{equation}\label{H_x7}
    H= H_i' - \cfrac{u_T^{(max)}}{x_7}\sqrt{H_r^2+H_{\theta}^2+H_h^2}
\end{equation}
\noindent where $H_i'$ is the control-independent term. The formulas for $H_i'$, $H_{r}$, $H_{{\theta}}$, and $H_{h}$, derived through extensive computations, are all linear in $\bm{\lambda}$
\begin{align}
    H_i' &= \lambda_{6}  \sqrt{\frac{\mu}{x_1^3}} \left( 1 + x_2  \cos{x_6} + x_3  \sin{x_6} \right)^2\\
    H_{{r}} &= \sqrt{\frac{x_1}{\mu}}  \left( \lambda_{2}  \sin{x_6} - \lambda_{3}  \cos{x_6} \right)\\
    \begin{split}
        H_{{\theta}} &= \frac{\sqrt{\frac{x_1}{\mu}}}{1 + x_2  \cos{x_6} + x_3  \sin{x_6}}  \left\{ 2  x_1  \lambda_{1} + \right.\\
        &+ \left[x_3 + \cos{x_6}  \left(2 + x_2  \cos{x_6} + x_3  \sin{x_6} \right) \right]  \lambda_{2} +\\
        &+ \left. \left[ x_3 + \sin{x_6}  \left(2 + x_2  \cos{x_6} + x_3  \sin{x_6} \right) \right]  \lambda_{3} \right\}
    \end{split}\\
    \begin{split}
        H_{{h}} &= \frac{\sqrt{\frac{x_1}{\mu}}}{2  \left( 1 + x_2  \cos{x_6} + x_3  \sin{x_6} \right)}  \left[  2  \left( x_4  \sin{x_6} + \right. \right.\\
        &- \left. x_5  \cos{x_6} \right)  \left( \lambda_{6} + x_2  \lambda_{3} - x_3  \lambda_{2} \right) +\\
        &+ \left. \left( x_4^2 + x_5^2 + 1 \right)  \left( \cos{x_6}  \lambda_{4} + \sin{x_6}  \lambda_{5} \right) \right].
    \end{split}
\end{align}

\noindent Linearity in $\bm \lambda$ implies that the costate equations are also homogeneous in $\bm \lambda$. As a result, if $\bm{\lambda}^*$ represents the optimal adjoint vector, then the optimal solution is associated with any $\bm \lambda$ such that ${\bm \lambda} = k_{\lambda} \, \bm{\lambda}^*$ (where $k_{\lambda}$ is an arbitrary positive constant).
% This leads to ${H} = k_{\lambda} H^*$ and consequently results in the expression:
% \begin{equation}\label{CostateHomogeneity}
%     \dot{{\boldsymbol{\lambda}}} = \frac{\partial {H}}{\partial \boldsymbol{x}} = k_{\lambda}  \frac{\partial H^*}{\partial \boldsymbol{x}} = k_{\lambda}  \dot{\boldsymbol{\lambda}^*}
% \end{equation}
Consequently, although the costate components have an unconstrained range, the adjoints at the initial time can be sought in an arbitrary interval $\left[\lambda_{min} \, \, \, \lambda_{max}\right]$, provided that $\lambda_{min}<0$ and $\lambda_{max}>0$. This interval can be set to $\left[-1, \, \, \, 1\right]$. This property helps the numerical search and is referred to as the scalability of the adjoints.

\indent Because $K_J$ ($>$ 0) is arbitrary (cf. Eq. (\ref{costfunc})), the boundary condition on the Hamiltonian at the final time in Eq. (\ref{NCgeneral})  can be ignored as an equality constraint \cite{pontanibook} and only the inequality
\begin{equation}\label{HF}
    H_f < 0
\end{equation}
must be satisfied. Concerning the transversality condition on the Hamiltonian at the initial time, it can be made explicit,
\begin{equation}\label{H0}
    \begin{split}
    H_0 = &\lambda_{10}\cfrac{\partial p(t_0)}{\partial t_0} +  \lambda_{20} \cfrac{\partial l(t_0)}{\partial t_0} +\lambda_{30}\cfrac{\partial m(t_0)}{\partial t_0} +  \\
    & +\lambda_{40}\cfrac{\partial n(t_0)}{\partial t_0} \, + 
      \lambda_{50}\cfrac{\partial s(t_0)}{\partial t_0}+\lambda_{60} \cfrac{\partial q(t_0)}{\partial t_0} - K_J 
\end{split}
\end{equation}
\noindent and can be rewritten as
\begin{equation}\label{H0_trans}
    H_0^{(ignition)}= H_0^{(coast)} - K_J
\end{equation}
\noindent with 
\begin{equation}
\begin{split}
     H_0^{(coast)} = &\lambda_{10}\cfrac{\partial p(t_0)}{\partial t_0} +  \lambda_{20} \cfrac{\partial l(t_0)}{\partial t_0} +\lambda_{30}\cfrac{\partial m(t_0)}{\partial t_0} +\\  & +\lambda_{40}\cfrac{\partial n(t_0)}{\partial t_0} \, + \lambda_{50}\cfrac{\partial s(t_0)}{\partial t_0}+\lambda_{60} \cfrac{\partial q(t_0)}{\partial t_0}
\end{split}
\end{equation}
\noindent where $H_0^{(coast)}$ represents the Hamiltonian function when no thrust is applied, whilst $H_0^{(ignition)}$ denotes the Hamiltonian function at $t_0$ when the propulsion system is ignited. Given the positivity of constant $K_J$, condition (\ref{H0_trans}) implies
\begin{equation}\label{H0_condition}
    H_0^{(ignition)} < H_0^{(coast)}.
\end{equation}
\noindent From the definitions of Hamiltonian Eq. (\ref{H_x7}), evaluated at the initial time, one obtains 
\begin{equation}\label{HFINAL}
    H_0^{(ignition)}= H_0^{(coast)} - \left[\cfrac{u_T^{(max)}}{x_7}\sqrt{H_r^2+H_{\theta}^2+H_h^2}\right].
\end{equation}
\noindent The term within brackets in (\ref{HFINAL}) is non-negative at all times, therefore relation (\ref{H0_condition}) turns out to be always satisfied.\\
\indent The boundary conditions for the adjoints at the initial time do not provide any useful information because those adjoints correspond to prescribed values of the state once $t_0$ is specified. Concerning the boundary conditions for the adjoint at the final time, they depend on the final state boundary condition vector $\bm{\Psi}_f$ and are
\begin{equation}\label{lambdafinal}
   \begin{split}
       \lambda_{1f}&= \nu_8 \, \, \, \,  \,\, \,\,\, \, \lambda_{2f}=  2 \, x_{2f} \nu_9 \, \, \, \, \,\, \, \,\, \, \lambda_{3f}= 2 \, x_{3f} \nu_9 \\
\lambda_{4f}&= 2 \, x_{4f} \nu_{10} \, \, \, \, \,\, \, \lambda_{5f}=  2 \, x_{5f} \nu_{10}\, \, \, \, \, \,\, \, \lambda_{6f}= 0.
     \end{split}
\end{equation}
\noindent The boundary condition on $\lambda_{1}$ yields an unknown time-invariant Lagrange multiplier, providing no beneficial information. A combination of the remaining relations leads to
\begin{equation}\label{adjbound2}
    \begin{split}
        \lambda_{2f}{x_{3f}} - \lambda_{3f}{x_{2f}}&=0\\
        \lambda_{4f}{x_{5f}} - \lambda_{5f}{x_{4f}}&=0\\
         \lambda_{6f}&=0.
    \end{split}
\end{equation}
\indent The minimum-time optimal control problem is posed as a well-defined Two-Point Boundary-Value Problem. The $8 \times 1$ parameter vector to optimize
includes the initial and final times of the transfer trajectory and the initial values of
the adjoints
\begin{equation}\label{Xvec}
    \bm{X} = \begin{bmatrix}
      t_0 & t_f & \lambda_{10} & \lambda_{20} & \lambda_{30} & \lambda_{40} & \lambda_{50} & \lambda_{60} \\ 
    \end{bmatrix}^T. 
\end{equation}
\noindent The corresponding $6 \times 1$ constraint vector includes the final boundary conditions on the state (\ref{finalbound2}) and the costate (\ref{adjbound2}). 
\begin{equation}\label{equalirycontr}
    \bm{Y} = \begin{bmatrix}
       x_{1f}-p_d \\ 
       x_{2f}^2+x_{3f}^2-e_d^2\\
       x_{4f}^2+x_{5f}^2-\tan^2{\cfrac{i_d}{2}} \\ \lambda_{2f}{x_{3f}} - \lambda_{3f}{x_{2f}} \\  \lambda_{4f}{x_{5f}} - \lambda_{5f}{x_{4f}} \\ \lambda_{6f} 
    \end{bmatrix} = \bm{0}.
\end{equation}
\noindent The two remaining constraints are the transversality conditions of the Hamiltonian at the final and initial time (\ref{HF}, \ref{H0_condition}).

\subsection{Method of Solution}
As a preliminary step for the numerical solution process, canonical units are introduced. The Distance Unit $\rm DU$ equals the mean equatorial radius of the main attracting body (the Moon), whereas the Time Unit $\rm TU$ is such that the lunar gravitational parameter is  $\mu = 1 \, \, {\rm DU^3}/{\rm TU^2}$.\\
\indent The indirect heuristic algorithm is based on the joint use of a heuristic technique (Particle Swarm Optimizer, PSO \cite{kennedyEbe}, in this work) and the analytical conditions of the optimal control problem. 
 The numerical solution process utilizes a population of individuals, where each individual corresponds to a possible solution. The fitness function is defined in the next subsection. Based on the necessary conditions, the solution technique consists of the following steps: 

 \begin{enumerate}
     \item Identify the known initial values of the state and costate variables and the minimal set of unknown values.
     \item For each individual $i$, with $i \leq N$ ($N$ being the total number of individuals), iterate the following sub-steps:
\begin{enumerate}
    \item select the unknown values for the initial time $t_0$, the final time $t_f$ and the costate components belonging to the minimal set identified at Step 1; 
    \item after selecting $t_0$, calculate the initial values of the state components;
    \item integrate numerically the state and costate equations while using Eqs. (\ref{anglesalfa})-(\ref{anglesbeta}) until $t = t_f$;
    \item evaluate the Hamiltonian at the final time. If $H_f < 0$ then go to sub-step (e), otherwise set the fitness function to infinity and go on to the next individual;
    \item evaluate the violations of the final conditions and the necessary conditions, then compute the fitness function, which is a measure of the final conditions violations. Then, go to the next individual.
\end{enumerate}

     \item Once the fitness functions are obtained for all the individuals that compose the population, use the PSO algorithm to update the position of individuals in the search space.
     \item Repeat all the steps until the fitness function of the best individual reaches a value lower than a threshold value or until the maximum number of iterations is reached.
     
 \end{enumerate}

\indent The PSO algorithm can be ended by imposing a fixed number of iterations or a condition that stops it when the global best has not changed for several stalled iterations. In this work, the $particleswarm$ routine provided by MATLAB is adopted with the use of $N=100$ particles for the numerical search. \\
\indent The $8 \times 1$ design parameters vector to be optimized is the same as in Eq. (\ref{Xvec}), except for the final time $t_f$, which is replaced by the time-of-flight $\Delta t$, to facilitate the computation. The lower and upper bounds that characterize the PSO search space for the parameter vector are
\begin{equation}
    \bm{LB} = \begin{bmatrix}
      t_{0_{I}} \\
      \Delta t_{min}\\
      \lambda_{min} \\
      \lambda_{min} \\
      \lambda_{min} \\
      \lambda_{min} \\
      \lambda_{min} \\
      \lambda_{min} 
    \end{bmatrix} \, \, 
    \bm{UB} = \begin{bmatrix}
      t_{0_{F}} \\
      \Delta t_{max} \\
      \lambda_{max} \\
      \lambda_{max} \\
      \lambda_{max} \\
      \lambda_{max} \\
      \lambda_{max} \\
      \lambda_{max} 
    \end{bmatrix}
\end{equation}

% \begin{equation}
% \begin{split}
%     \bm{LB} &= \begin{bmatrix}
%       t_{0_{I}} & 25 \, \, {\rm d} & -1 & -1 & -1 & -1 & -1 & -1 
%     \end{bmatrix} \\
%     \bm{UB} &= \begin{bmatrix}
%       t_{0_{F}} & 45 \, \, {\rm d} & 1 & 1 & 1 & 1 & 1 & 1 
%     \end{bmatrix}
%     \end{split}
% \end{equation}
\noindent where $t_{0_{I}}$ and $t_{0_{F}}$ are the initial and final reference epoch of a specific period of the Lunar Gateway, extracted from Gateway's ephemeris \cite{lee2019white}, whereas $\lambda_{min}=-1$ and  $\lambda_{max}=1$. Furthermore, $\Delta t_{min}$ and $\Delta t_{max}$ are set to 25 and 45 days, respectively. The search interval of the flight time was chosen in accordance with the approximate analytical solution (cf. Appendix A). 
\subsection{Numerical results}
The point-mass spacecraft is characterized by the following propulsion parameters:
\begin{equation}\label{proppar}
        u_T^{(max)}  = 4.903 \cdot 10^{-7} \, \, \rm{\cfrac{km}{s^2}} \,\, \, \, \, 
        c = 30 \, \, \rm{\cfrac{km}{s}}.
\end{equation}
These parameters, when applied to a vehicle with a mass equivalent to the Automated Transfer Vehicle (ATV) of 15000 kg \cite{amadieu1999automated} result in a maximum thrust magnitude of $T^{(max)}=7.3545 \cdot 10^{-3}$ N.
The spacecraft of interest is placed on the Lunar Gateway orbit, and the initial time of the transfer (i.e., $t_0$) is selected along a single period of the departing orbit.
The final orbit is a LLO with the following desired orbit elements: 
\begin{equation}\label{edid}
p_d = R_M + 100 \, \, {\rm{km}} \, \, \, \, \,  e_d = 0 \, \, \, \,  \,  i_d = 90^{\circ}
\end{equation}
where $R_M = 1737.4 \, \, \rm{km}$ is the mean equatorial radius of the Moon.
The objective of the optimization algorithm is an equality constraint-inspired fitness function $\Tilde{J}$ defined as
\begin{equation}
    \resizebox{.88\hsize}{!}{$\Tilde{J} = \sqrt{w_1 \, Y_1^2\, + \, w_2 \, Y_2^2\, + \, w_3 \, Y_3^2\, + \, w_4 \, Y_4^2\, + \, w_5 \, Y_5^2\, + \, w_6 \, Y_6^2}$}
\end{equation}
\noindent where $Y_i$ (for $i = 1 \dots 6$) are the elements of the constraint vector in Eq. (\ref{equalirycontr}), and the $w_i$ are weights specifically chosen to give comparable relevance to the boundary conditions to satisfy. After a tuning phase, the  weights are
\begin{equation}
    w_1=w_3=w_4=w_5=w_6 = 1 \, \, \,w_2 = 100.
\end{equation}

\noindent The decision to assign a higher weight to $w_2$ is justified by the fact that convergence for the eccentricity turned out to be more challenging.
\noindent PSO ended with a fitness function value of $3.2032 \cdot 10^{-2}$. To obtain a more accurate solution, the \textit{fminsearch} MATLAB routine is employed, resulting in $\Tilde{J} = 3.0863 \cdot 10^{-10}$. \noindent The converged values for the design parameters appear in Table \ref{tabsol}.
\bgroup
\def\arraystretch{1.2}
\begin{table}[h] 
\centering
\begin{tabular}{cc}
\toprule
\text{Parameter} & \text{Solution} \\
\midrule
$t_0$&774909.503 TU\\ 
$\Delta t$&3027.130 TU\\ 
$\lambda_{10}$&0.0209 \\
$\lambda_{20}$&1 \\
$\lambda_{30}$&0.2947 \\
$\lambda_{40}$&0.0623\\
$\lambda_{50}$&-0.0161\\
$\lambda_{60}$&0.0148\\
\bottomrule
\end{tabular}
\vspace*{3mm}
\caption{Solution of the TPBVP}
\label{tabsol}
\end{table}
\egroup

\indent The optimal starting time is May 25th 2025 at 14:06:08 UTC and the time of flight equals $ \Delta t = 36 \, \, {\rm d} \, \, 5 \, \, {\rm hrs} \, \, 40 \, \, {\rm min} \, \, 24 \, \, {\rm s} $, whereas the final mass ratio is $0.9488$. Table \ref{tabstate} provides the spacecraft's orbit elements at the initial and final time. The orbital elements of the spacecraft at the initial time are the osculating parameters of Gateway at the beginning of the transfer. 

\bgroup
\def\arraystretch{1.2}
\begin{table}[h] 
\centering
\begin{tabular}{ccc}
\toprule
\text{COE} & ${t_0}$ & ${t_f}$\\
\midrule
$a$ [km]&$2.327\cdot10^{4}$&$1.837\cdot10^{3}$\\ 
$e$&$0.866$ & $1.653\cdot10^{-6}$\\ 
$i$  &$97.916 \degree$ & $90.000 \degree$\\
$\Omega$  &$ -62.647 \degree$ &$-17.719 \degree$ \\
$\omega$  &$84.175 \degree$ & $88.297 \degree$ \\
$\theta_*$  &$166.97 \degree$ & $-38.025 \degree$  \\
\bottomrule
\end{tabular}
\vspace*{3mm}
\caption{Spacecraft initial and final orbit elements}
\label{tabstate}
\end{table}
\egroup

\begin{figure}[H]
   \includegraphics[width=1\columnwidth]{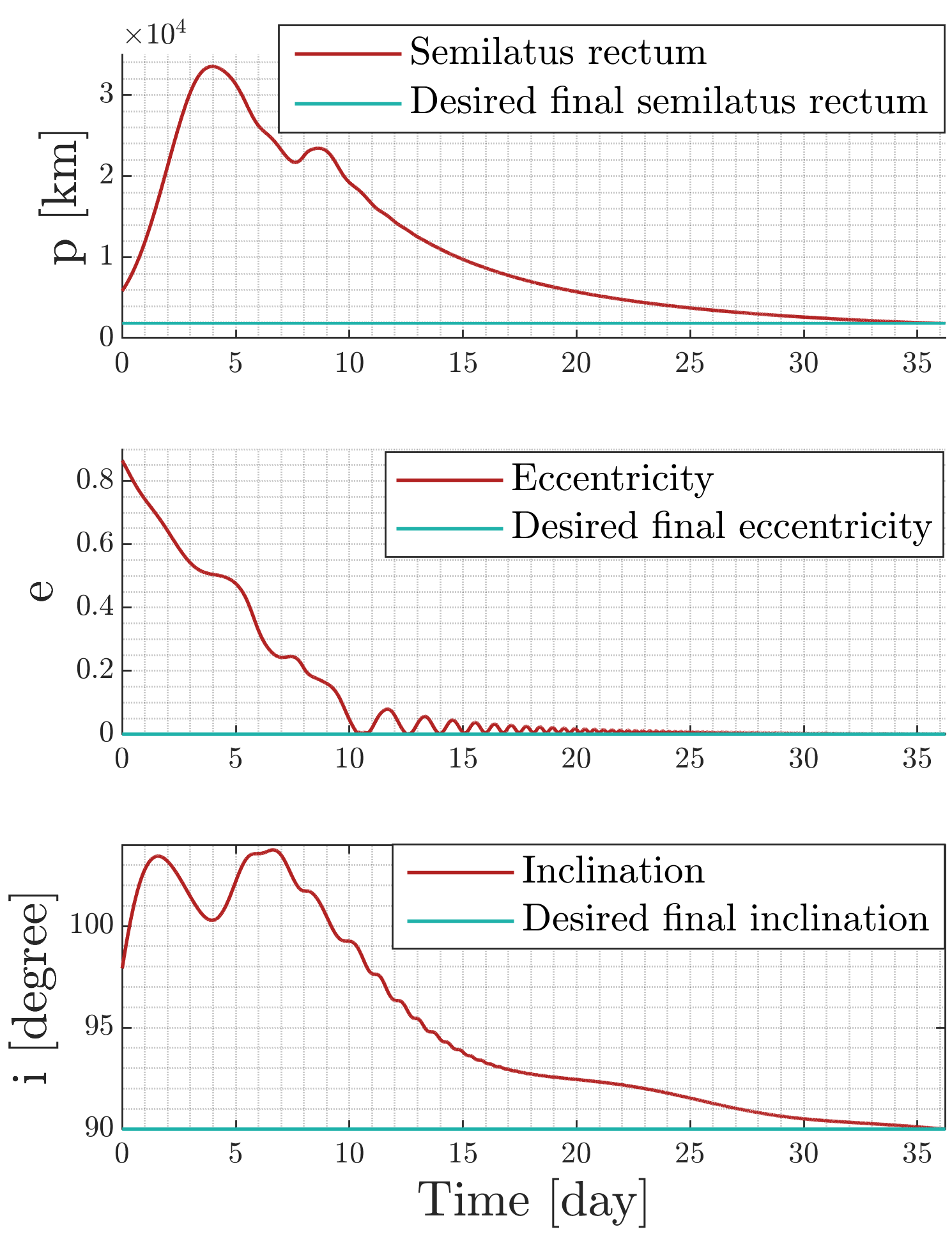}
   \caption{Time evolution of the controlled orbital parameters}
    \label{pei_opt}
\end{figure}

\begin{figure}[H]
     \includegraphics[width=\columnwidth]{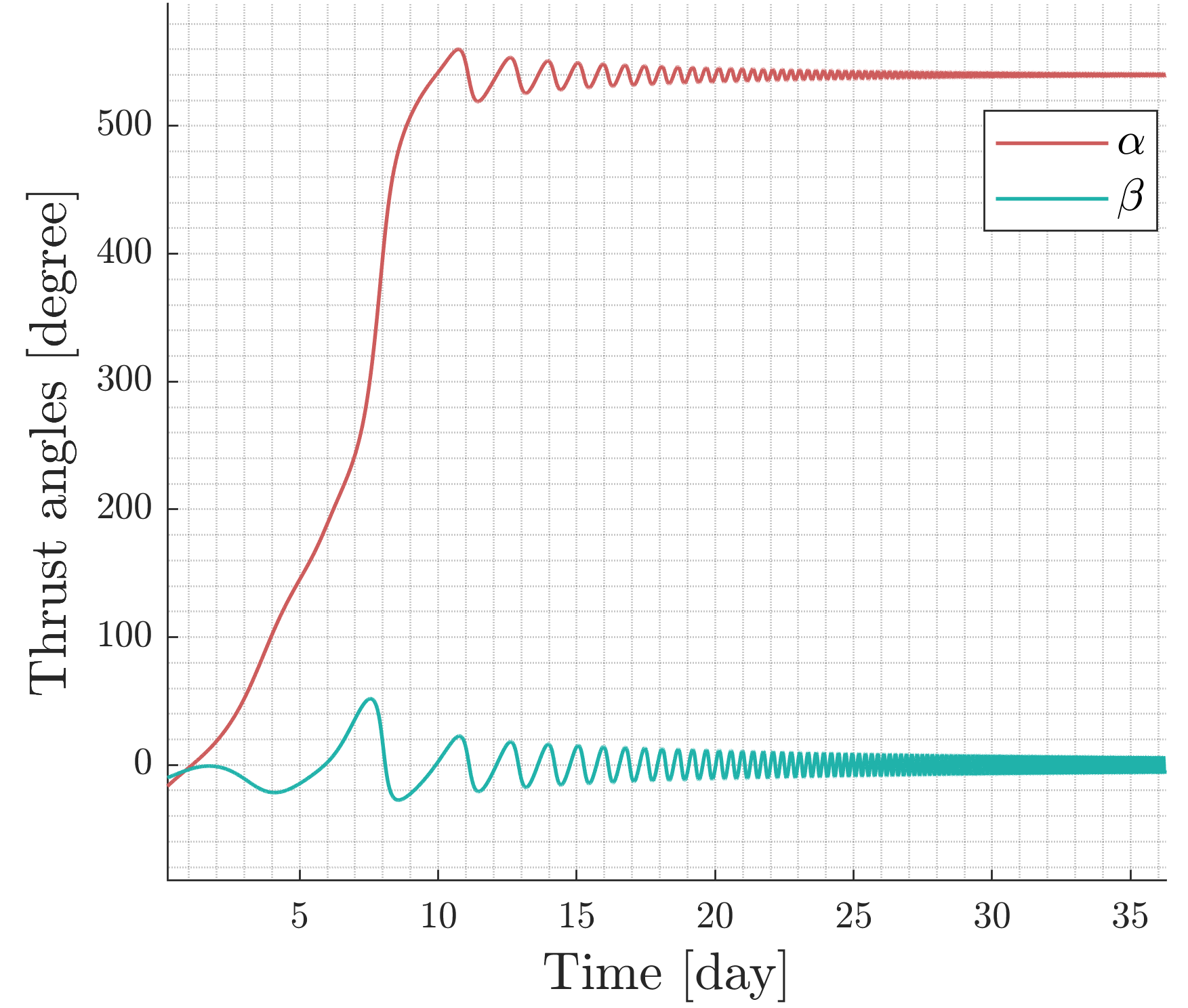}
    \caption{Time evolution of the thrust angles}
    \label{alfabeta_opt}
\end{figure}

\begin{figure}[H]
     \includegraphics[width=\columnwidth]{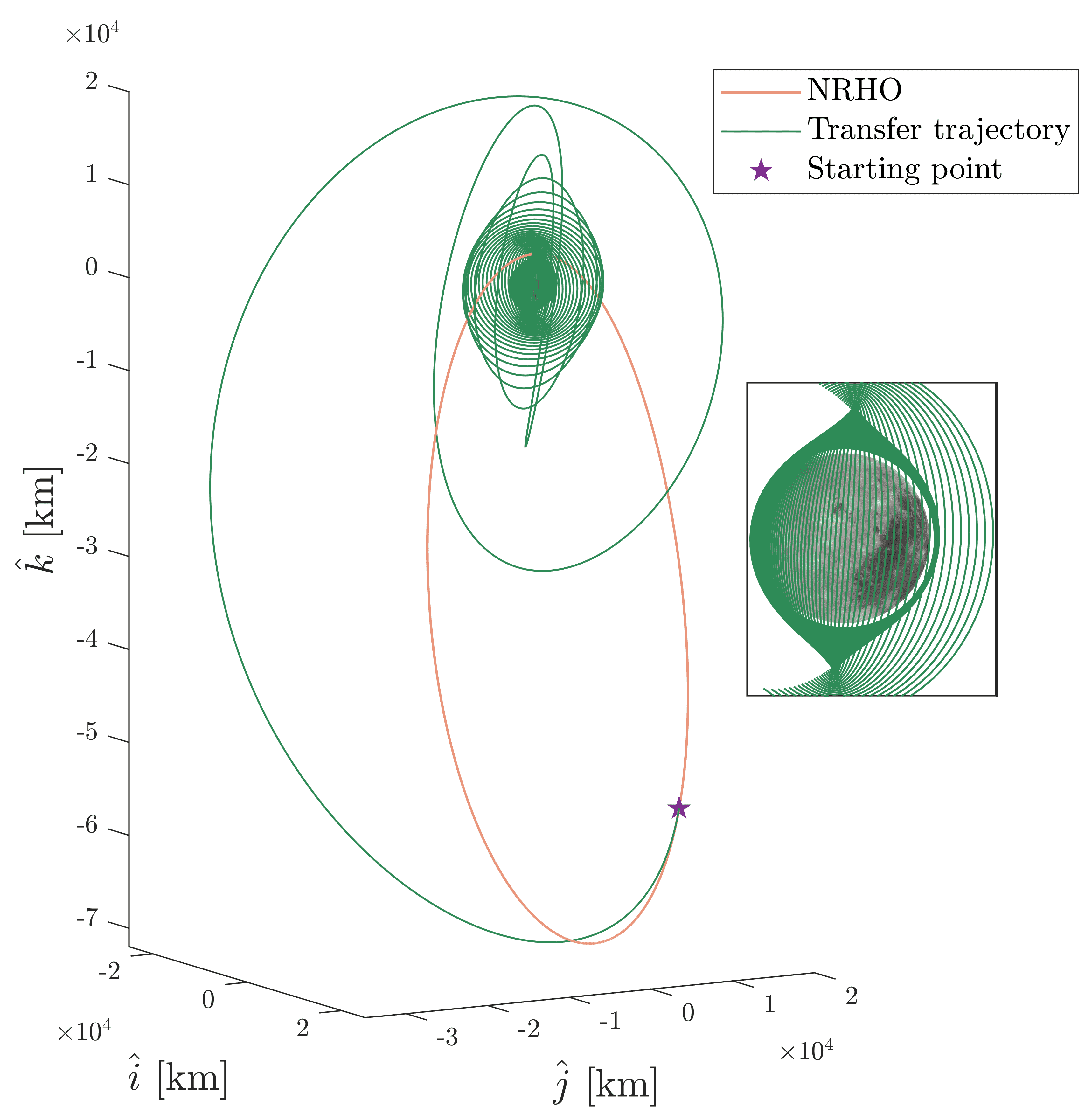}
    \caption{Optimal transfer trajectory in Moon-centered synodic frame}
    \label{trajopt}
\end{figure}

Figure \ref{pei_opt} depicts the time histories of $p$, $e$, and $i$, whereas Figs. \ref{alfabeta_opt} and \ref{trajopt} portray the optimal thrust angles and transfer trajectory.
%\begin{figure}[h]
 %  \includegraphics[width=0.92\columnwidth]{pvec_opt.png}
  %  \caption{Time evolution of the semilatus rectum}
   % \label{p_opt}
%\end{figure}
%\begin{figure}[h]
%    \includegraphics[width=0.92\columnwidth]{evec_opt.png}
 %   \caption{Time evolution of the eccentricity}
  %  \label{e_opt}
%\end{figure}
%\begin{figure}[H]
  %   \includegraphics[width=0.92\columnwidth]{ivec_opt.png}
    %\caption{Time evolution of the inclination}
    %\label{i_opt}
%\end{figure}
The time evolution of the semilatus rectum in Figure \ref{pei_opt} shows a first considerable increase, followed by a smooth decrease before converging to the desired value. This behavior is attributed to the plane change maneuver performed with low thrust. The time history of the eccentricity shows a rapid decrease in the first 10 days, from a high value, which is typical of the NRHOs, to a relatively low one, then reducing to 0 in the remaining days. From the time evolution of the inclination it is clear that during the transfer, the orbital plane changes continuously. \\
\indent Concerning the in-plane thrust angle $\alpha$, in order to get a more effective visualization in Fig. \ref{alfabeta_opt}, it is represented in a more suitable interval,
$-60 \degree \leq \alpha \leq 600 \degree$.
The out-of-plane angle $\beta$ is constrained to $-90 \degree \leq \beta \leq 90 \degree$.
Figure \ref{trajopt} represents the optimal transfer trajectory in the Moon-centered synodic frame. It is interesting to note that the starting point is close to the aposelenium of the NRHO and that the spacecraft travels many orbits
about the Moon before arriving at the final desired LLO. These behaviors
are due to the low-thrust parameters used along the minimum-time transfer.

\section{{Guidance, Attitude Control, and Actuation}}\label{sect4}
The request for an increasing autonomy of space missions has driven research in feedback orbit control. Real scenarios require considering the vehicle no longer as a point mass, but as a 6-degree-of-freedom system, and the thrust not exactly supplied in the desired direction at all times. To this end, nonlinear feedback control laws for both orbit and attitude control are derived through a Lyapunov-based approach. Attitude actuators and the related dynamics are modeled as well. The study that follows considers the overall spacecraft dynamics, with the intent of designing an effective guidance, control, and actuation architecture. 

\subsection{Nonlinear orbit control}
Previous research \cite{gurfilgf} demonstrated that any state associated with elliptic orbits is accessible when the spacecraft dynamics is subject to the Lagrange planetary equations for MEE. This embodies the theoretical basis for applying nonlinear techniques to orbital control.\\
\indent In practical mission scenarios, the final values of $\{a, \, e, \, i, \, \Omega, \, \omega  \}$ (or a subset of them) are specified. The desired operational conditions can be defined in terms of $\bm z$ only. The target state is properly defined by $ {\bm \Psi}({\bm{z}}) = \bm 0$, and is problem dependent. A feedback control law capable of driving the dynamical system toward the target conditions is 
\begin{equation}\label{fcl}
    {\bm u}_T = -{\bm u}_T^{(max)}\cfrac{x_7\left({\bm b}+ {\bm a}_P\right)}{max \left\{{\bm u}_T^{(max)}, \, x_7|{\bm b}+ {\bm a}_P|\right\}}
\end{equation}
\noindent where
\begin{equation}
    {\bm b} = {\bm G}^T\left(\cfrac{\partial {\bm \Psi}}{\partial {\bm z}}\right)^T{\bm K}^T{\bm \Psi}
\end{equation}
\noindent $\bm K$ being a symmetric and positive definite gain matrix.
This feedback law enjoys quasi-global stability \cite{pontanibook}. Moreover, Eq. (\ref{fcl}) incorporates the saturation condition on ${\bm u}_T$ (i.e. $|{\bm u}_T| \leq {u}_T^{(max)}$), and provides a control law that allows steerable and throttleable propulsive thrust, with time-varying magnitude and direction. It is worth pointing out that the Lyapunov method provides sufficient conditions for stability, meaning that violation of those conditions (in limited time intervals) does not necessarily imply instability. Further analysis is needed in order to identify the attracting set, which corresponds to ${\bm b} = \bm 0$. The last condition is problem-dependent and shall be investigated in a specific case. 

The subsequent discussion delves into nonlinear orbit control, aiming to guide the spacecraft toward a desired orbit with specified final values for semimajor axis, eccentricity, and inclination (represented as $a_d$, $e_d$, and $i_d$, respectively).
The target set corresponds to  
\begin{equation}\label{targesetreal}
    \bm{\Psi}=  \begin{bmatrix}
       x_{1}-p_d \\
       x_{2}^2+x_{3}^2-e_d^2 \\
       x_{4}^2+x_{5}^2-\tan^2{\cfrac{i_d}{2}}
    \end{bmatrix}= \bm{0}.
\end{equation}
\noindent To investigate the stability of the feedback law (\ref{fcl}), the components of the vector $\bm b$ are derived analytically 
\begin{equation}\label{belement}
    \begin{split}
        b_1 &= -2k_2 \sqrt{\cfrac{x_1}{\mu}}\left(x_3 \cos{x_6}-x_2\sin{x_6}\right)\Psi_2\\
         b_2 &= \cfrac{2}{\eta}\sqrt{\cfrac{x_1}{\mu}}\left[k_1x_1\Psi_1+k_2\Psi_2\left(\eta^2+x_2^2+x_3^2-1\right)\right]\\
         b_3 &= \cfrac{k_3}{\eta}\sqrt{\cfrac{x_1}{\mu}}\left(x_4 \cos{x_6} +x_5 \sin{x_6}\right)\Psi_3\left(x_4^2+x_5^2+1\right).
    \end{split}
\end{equation}
\noindent In (\ref{belement}) $\{k_1, \, k_2, \, k_3\}$ are the diagonal elements of $\bm K$. The attracting set gathers all the dynamical states such that $\bm b=0$, and includes the following five subsets \cite{pontanibook}:
\begin{enumerate}
    \item $x_1 = 0$, i.e., rectilinear trajectories;
    \item $x_1 = p_d$, $x_2^2+x_3^2=e_d^2$, $x_4=x_5=0$, i.e., equatorial orbits with semilatus rectum $p_d$ and eccentricity $e_d$;
    \item $x_1 = p_d$, $x_2^2+x_3^2=0$, $x_4=x_5=\tan^2{\cfrac{i_d}{2}}$, i.e., circular orbits with semilatus rectum $p_d$ and inclination $i_d$;
    \item $x_1 = p_d$, $x_2^2+x_3^2=0$, $x_4=x_5=0$, i.e., circular equatorial orbits with semilatus rectum $p_d$;
    \item $x_1 = p_d$, $x_2^2+x_3^2=e_d^2$, $x_4=x_5=\tan^2{\cfrac{i_d}{2}}$, i.e., the target set.
\end{enumerate}
Because $\bm \Psi$ is continuous and $\dot{V}<0$ (outside the attracting set, denoted with $A$), the condition $V({\bm z}) \leq V({\bm z}_0)$ identifies the compact set $C$. Furthermore, the invariant set must be sought in $A \cap C$; it is composed of all the dynamical states (in the attracting set) that remain unchanged when the overall non-Keplerian acceleration vector is the null vector (${\bm a} \equiv 0$). Once the invariant set is reached, ${\bm b} \equiv 0$, which implies $\dot{\bm b} \equiv 0$ while ${\bm a} \equiv 0$. The time derivatives of the three components of ${\bm b}$ for the specific case at hand are
\begingroup
\thinmuskip=0mu
\medmuskip=0mu
\thickmuskip=0mu
\begin{equation}\label{bdot}
    \resizebox{.88\hsize}{!}{$\begin{split}
        \dot{b}_1 &= \frac{2 k_2}{x_1}\left(x_2 c_{x_6}+x_3 s_{x_6}\right)\Psi_2\\
        \dot{b}_2 &= \frac{2}{x_1}\left(x_2 s_{x_6}-x_3 c_{x_6}\right)\left[k_1 x_1\Psi_1+ k_2 \Psi_2\left(-\eta^2+x_2^2+x_3^2-1\right)\right]\\
        \dot{b}_3 &=\frac{k_3}{x_1}\Psi_3\left(x_4^2+x_5^2+1\right)\left(x_2x_5-x_3x_4+x_5 c_{x_6}-x_4 s_{x_6}\right).
    \end{split}$}
\end{equation}
\endgroup
\noindent It is clear that subset 1 does not belong to the invariant set. On the contrary, the other four subsets form the invariant set. Yet, convergence toward subsets 2, 3 and 4 is only theoretical. In fact, 
%the Lyapunov function (\ref{lyap}) can be rewritten in terms of orbit elements as 
% \begingroup
% \thinmuskip=0mu
% \medmuskip=0mu
% \thickmuskip=0mu
% \begin{equation}\label{ly2}
%      \resizebox{.88\hsize}{!}{$V = \frac{1}{2}\left[k_1\left(p-p_d\right)^2+k_2\left(e^2-e_d^2\right)^2+k_3\left(\tan^2{\frac{i}{2}}-\tan^2{\frac{i_d}{2}}\right)^2  \right]$}
% \end{equation}
% \endgroup
% \noindent Since all the subsets are characterized by the condition $\Psi_1 =0$, the Lyapunov function (\ref{ly2}) can be evaluated at $p=p_d$ and, as a result, it can be regarded as a function of two variables (i.e. $e$ and $i$). In order to investigate the equilibrium conditions the partial derivatives of $V$, with respect to the eccentricity and the inclination, are retrieved
% \begingroup
% \thinmuskip=0mu
% \medmuskip=0mu
% \thickmuskip=0mu
% \begin{equation}\label{Vpartialder}
%  \resizebox{.88\hsize}{!}{$\begin{split}
%     &\frac{\partial V}{\partial e} = 2 \, k_2 \, e\left(e^2-e_d^2\right)\\
%     &\frac{\partial V}{\partial i} =2 \, k_3 \tan{\frac{i}{2}}\left(1+\tan^2{\frac{i}{2}}\right)\left(\tan^2{\frac{i}{2}}-\tan^2{\frac{i_d}{2}}\right)\\
%     &\frac{\partial ^2 V}{\partial e^2}=2 \, k_2\left(3e^2-e_d^2\right)\\
%     &\frac{\partial ^2 V}{\partial i^2} =-\frac{k_3}{2}\left(1+\tan^2{\frac{i}{2}}\right)\left[3\tan^2{\frac{i}{2}}\left(\tan^2{\frac{i_d}{2}}-1\right)-5\tan^4{\frac{i}{2}}+\tan^2{\frac{i_d}{2}}\right]\\
%     &\frac{\partial ^2 V}{\partial e \partial i } =0
%     \end{split}$}
% \end{equation}
% \endgroup
one can prove that subsets 2 and 3 are associated with saddle points, while subset 4 corresponds to a local maximum of the Lyapunov function. Subset 5 (i.e. the target set) is related to the global minimum, and is the only stable equilibrium condition \cite{pontanibook}. The dynamical system of interest is characterized by global (numerical) convergence toward the desired operational conditions.

\subsection{Commanded attitude}\label{commandedsection}
The body is no longer regarded as a point mass, instead, it is modeled as a rigid body characterized by an inertia matrix ${\bm J}_c^{(B)}$ and attitude dynamics described by Eqs. (\ref{qpunto}) and (\ref{eulerdef}). \\
\indent The space vehicle is equipped with an active attitude control system, to track a time-varying desired attitude. The vehicle must point its ${\hat{b}_{1}}$ axis (aligned with the main thruster) toward the direction of the desired thrust. This means that the unit vector ${\hat{b}_{c_1}}$ is aligned with the commanded thrust acceleration. For the orientation of the other two commanded axes ${\hat{b}_{c_2}}$ and ${\hat{b}_{c_3}}$, sampling times $[t_0,\, \dots, \, t_N]$ are defined, e.g., the times at which each guidance interval begins,
\begin{itemize}
    \item In $[t_0, \, t_1]$
       \begin{equation*} 
        \begin{split}
            \hat{b}_{c_2}(t) &= \cfrac{\hat{c}_3^M \times  \hat{b}_{c_1}(t)}{|\hat{c}_3^M \times  \hat{b}_{c_1}(t)|}\\
            \hat{b}_{c_3}(t) &=  \hat{b}_{c_1}(t) \times  \hat{b}_{c_2}(t)
        \end{split}
        \end{equation*}
    \item In $[t_1, \, t_2]$
       \begin{equation*} 
        \begin{split}
            \hat{b}_{c_2}(t) &= \cfrac{\hat{b}_{c_3}(t_1) \times  \hat{b}_{c_1}(t)}{|\hat{b}_{c_3}(t_1) \times  \hat{b}_{c_1}(t)|}\\
            \hat{b}_{c_3}(t) &=  \hat{b}_{c_1}(t) \times  \hat{b}_{c_2}(t)
        \end{split}
        \end{equation*}
    \item In $[t_k, \, t_{k+1}]$
        \begin{equation*} 
        \begin{split}
            \hat{b}_{c_2}(t) &= \cfrac{\hat{b}_{c_3}(t_k) \times  \hat{b}_{c_1}(t)}{|\hat{b}_{c_3}(t_k) \times  \hat{b}_{c_1}(t)|}\\
            \hat{b}_{c_3}(t) &=  \hat{b}_{c_1}(t) \times  \hat{b}_{c_2}(t).
        \end{split}
        \end{equation*}
\end{itemize}
\noindent This method avoids discontinuities in the definition of $\hat{b}_{c_2}(t)$, provided that the sampling time $\Delta t = t_{k+1}-t_k$ is sufficiently short (in relation to the thrust angles oscillations). Hence, once the time history of the control thrust acceleration is known, the commanded body frame is identified,
\begin{equation}\label{commbodyf}
      \underline{\underline{C}} = [{\hat{b}_{c_1}} \: \: {\hat{b}_{c_2}} \: \: {\hat{b}_{c_3}}].
 \end{equation}

\subsection{Nonlinear Attitude Control}
For the scenario at hand, a triaxial attitude control law is used, with reference to the desired attitude described in section \ref{commandedsection}. In the attitude tracking problem, the spacecraft attitude and angular rate must track known commanded time histories (i.e. drive the actual body frame $\underline{\underline{B}}$ toward the commanded frame $\underline{\underline{C}}$),
\begin{equation}
    \begin{split}
          \{q_{0}(t), \, {\bm{q}{(t)}}\} & \longrightarrow  \{q_{c0}(t), \, {\bm{q}_c(t)}\}\\
        {\bm{\omega}(t)} & \longrightarrow {\bm{\omega}_c(t)}
    \end{split}
\end{equation}
\noindent where $ {\bm{\omega}(t)}$ and ${\bm{\omega}_c(t)}$ include the components of the angular velocity and the commanded angular velocity in the respective frames, respectively. \\
\indent The control torque ${\bm{T}}_c$ that enable this must be sought. To do so, the error quaternions are introduced. The error quaternions $\{q_{e0}, \, {\bm{q}}_e\}$ are a set of quaternions related to the rotation matrix $\underset{\rm B \leftarrow C}{\rm \textbf{R}}$ that links $\underline{\underline{C}}$ to $\underline{\underline{B}}$.
The error quaternion is given by \cite{pontanibook}

\begin{equation}
    \begin{bmatrix}
        q_{e0} \\
        {\bm{q}}_e
    \end{bmatrix} =  \begin{bmatrix}
        q_{c0} & {\bm{q}}_c^T\\
        {\bm{q}}_c & q_{c0}{\bm I}_{3 \times 3}-\Tilde{{\bm{q}}}_c
    \end{bmatrix} 
    \begin{bmatrix}
         q_{0} \\
        {\bm{q}}.
    \end{bmatrix}
\end{equation}
\noindent The feedback law must drive $\underset{\rm B \leftarrow C}{\rm \textbf{R}}$ to the identity matrix, i.e. $q_{e0} = \pm 1, \, {\bm{q}}_e = \bm 0$. Using the Lyapunov theorem and the LaSalle principle, the following feedback law, which enjoys quasi-global stability properties, can be derived \cite{pontanibook}: 
\begin{equation}\label{controlTcfinal}
\begin{split}
     {\bm T}_c =& \bm{\tilde{\omega}}\bm{J}_c^{(B)}\bm{{\omega}}-\bm{M}_c + \bm{J}_c^{(B)} \dot{\bm \omega}_c\\
 &-\bm{J}_c^{(B)} \, {\bm A}^{-1} \, \bm B \, {\bm \omega}_D- {\rm sign}\{q_{e0}(t_0)\}\bm{J}_c^{(B)} \, {\bm A}^{-1} \, {\bm q}_e
 \end{split}
\end{equation}
\noindent where $\dot{\bm \omega}_c$ is a known function of time because ${\bm \omega}_c(t)$ is specified, and ${\bm \omega}_D={\bm \omega}-{\bm \omega}_c$. Matrices $\bm A$ and $\bm B$ are constant and positive definite gain matrices that affect the convergence speed, therefore they must be accurately selected through numerical simulations. Matrix $\bm A$ is also symmetric.

\subsection{Gain selection}
In order to achieve convergence to the desired conditions, it is essential to determine the appropriate gain matrices for the Lyapunov-based controllers (denoted with $\bm K$ for the orbit control law and $\bm A$ and $\bm B$ for the attitude feedback law), in other words, the gains of the control laws need to be tuned. \\
\indent Concerning the guidance,  the symmetric and positive definite gain matrix is selected to be diagonal and its elements are carefully tuned to ensure that the feedback solution closely resembles the optimal solution in terms of flight time. \\
\indent Regarding the triaxial attitude control law, the two gain matrices must be tuned to successfully track the commanded attitude while avoiding the commanded torque reaching excessive values, not compatible with the technological limitations of the actuation devices (reaction wheels, in this study).\\
\indent The process of tuning the control gains can be time-consuming if performed by trial-and-error, hence after a first rough choice of the orders of magnitude, a numerical optimization algorithm was employed. Specifically, the $fminsearch$ MATLAB routine was utilized to optimize the gain matrix elements. This routine efficiently explores the parameter space to find suitable values, associated with the desired control response.  

\subsection{Steering law for spacecraft equipped with reaction wheels}\label{sectionwheel}

The nonlinear feedback law designates the commanded torque needed for the spacecraft to track the desired attitude. The dynamics of the actuators must be taken into account, to compute the actual torque that they can exert. Steering algorithms are indeed needed in order to define the motion of multiple momentum exchange devices in an organized, coordinated way, so that the desired control is obtained. \\
\indent From the time history of the thrust angles of the transfer trajectory at hand (cf. Fig. \ref{alfabeta_opt}), it is expected that no agile and fast slews are necessary to the spacecraft to track the desired attitude. Hence, an array of reaction wheels is chosen as an adequate architecture to address the attitude-tracking problem.\\
\indent The steering logic when arrays of three or more reaction wheels are used encounters no singularity if the axes of the wheels are not coplanar. The actual torque supplied by the reaction wheels is \cite{pontanibook}
\begin{equation}
     \bm{T}_a = -\Tilde{\bm \omega} {\bm W} \, {\bm \omega}_{s} - {\bm W} \, \dot{\bm \omega}_{s}
\end{equation}
where
\begin{equation}
    {\bm \omega}_s = \begin{bmatrix}
        \omega_{s1} & \dots & \omega_{sN}
    \end{bmatrix}^T
\end{equation}
\noindent $\omega_{si}$ denotes the wheel's angular velocity with respect to the spacecraft's body frame $\underline{\underline{B}}$, whereas 
\begin{equation}\label{actmat}
    \bm W := \begin{bmatrix}
        I_{s1}{\bm a}_1 & \dots & I_{sN}{\bm a}_N
    \end{bmatrix}
\end{equation}
\noindent is termed the actuation matrix, $I_{si}$ is the inertia moment of the $i^{th}$ wheel along its spin axis ${\bm a}_i$ (where ${\bm a}_i$ is a 3x1 vector that includes the 3 components of the axis of wheel $i$ along the spacecraft body axes). This matrix has maximum rank because ${\bm a}_1 \, \dots \, {\bm a}_N$ are constant and not coplanar. 
The values of $\dot{\bm \omega}_{s}$ can be found by applying the following simplifying approximation. Since the spacecraft has generally low values of $|\bm \omega|$ one can assume that $|{\bm W} \, \dot{\bm \omega}_{s}| >> |\Tilde{\bm \omega} {\bm W} \, {\bm \omega}_{s}|$. Thus, the control torque is simplified, 
\begin{equation}
   \bm{T}_c = -{\bm W} \, \dot{\bm \omega}_{s}.
\end{equation} 
\noindent Therefore, for a given commanded torque ${\bm T}_c$, it is desirable to obtain the values of $\dot{\bm \omega}_{s}$ that yield the desired control torque. The values of the angular accelerations of the wheels are found using the Moore-Penrose pseudoinverse law, which leads to 
\begin{equation}
\dot{\bm \omega}_{s} =-{\bm W} \left({\bm W}{\bm W}^T\right)^{-1} {\bm T}_c.
\end{equation}
\noindent It is worth emphasizing that $\left({\bm W}{\bm W}^T\right)$ is never singular, provided that at least 3 directions of the ${\bm a}_i$ are non-coplanar.

\subsection{Guidance, Control, and Actuation architecture}

The feedback algorithm architecture involves two control loops, an outer loop and an inner one. The outer loop deals with orbit guidance and attitude control while the inner loop tackles the actuators. The guidance, control, and actuation architecture has the final purpose of identifying the wheels' angular accelerations, and includes the following 6 steps, to complete in each sampling interval:

\begin{enumerate}
    \item Numerical trajectory propagation is performed using nonlinear orbit control.
    \item The thrust direction determines the commanded body frame $\underline{\underline{C}}$.
    \item The commanded rotation matrix is defined, and relates the MCI frame to the commanded body frame, $\underset{\rm C \leftarrow MCI}{\rm \textbf{R}}$.
    \item From $\underset{\rm C \leftarrow MCI}{\rm \textbf{R}}$ the commanded attitude is retrieved in terms of quaternions, and the commanded angular velocity and its derivative are obtained.
    \item The nonlinear attitude control system calculates the commanded torque ($\bm{T}_c$).
    \item Steering algorithms translate the torque into individual reaction wheels’ rate commands.
\end{enumerate}

\noindent The overall dynamics in the sampling interval is simulated through the following three steps:

\begin{enumerate}[label=\Alph*.]
    \item The actual torque ($\bm{T}_a$) is computed, considering the actuators' dynamics and technological constraints.
    \item The actual torque is applied to the dynamical system and yields the actual spacecraft attitude (body frame $\underline{\underline{B}}$) and the actual thrust direction.
    \item  Orbit dynamics is propagated with the use of the actual thrust direction, up to the next sampling time.
\end{enumerate}

\indent The attitude actuation system is represented by an array of reaction wheels organized in a proper geometry. For triaxial control, at least 3 non-aligned reaction wheels are needed. Due to redundancy requirements, this work uses a pyramidal array composed of 4 reaction wheels. The configuration of the four reaction wheels composing the array is assigned by their spin axes,
\begin{equation}
\begin{split}
    {\hat{a}}_1 &= \begin{bmatrix}
        \cfrac{1}{\sqrt{3}} & \cfrac{1}{\sqrt{3}} & \cfrac{1}{\sqrt{3}}
    \end{bmatrix}\,
     \underline{\underline{B}}^T \\
     {\hat{a}}_2 &= \begin{bmatrix}
        \cfrac{1}{\sqrt{3}} & -\cfrac{1}{\sqrt{3}} & \cfrac{1}{\sqrt{3}}
    \end{bmatrix}\,\, 
     \underline{\underline{B}}^T\\
      {\hat{a}}_3 &= \begin{bmatrix}
        -\cfrac{1}{\sqrt{3}} & \cfrac{1}{\sqrt{3}} & \cfrac{1}{\sqrt{3}}
    \end{bmatrix}\,\, 
     \underline{\underline{B}}^T \\
     {\hat{a}}_4 &= \begin{bmatrix}
        -\cfrac{1}{\sqrt{3}} & -\cfrac{1}{\sqrt{3}} & \cfrac{1}{\sqrt{3}}
    \end{bmatrix}\,\, 
     \underline{\underline{B}}^T.
\end{split}
\end{equation}

\subsection{Numerical results in nominal flight conditions}
\begin{figure}[H]
     \includegraphics[width=0.97\columnwidth]{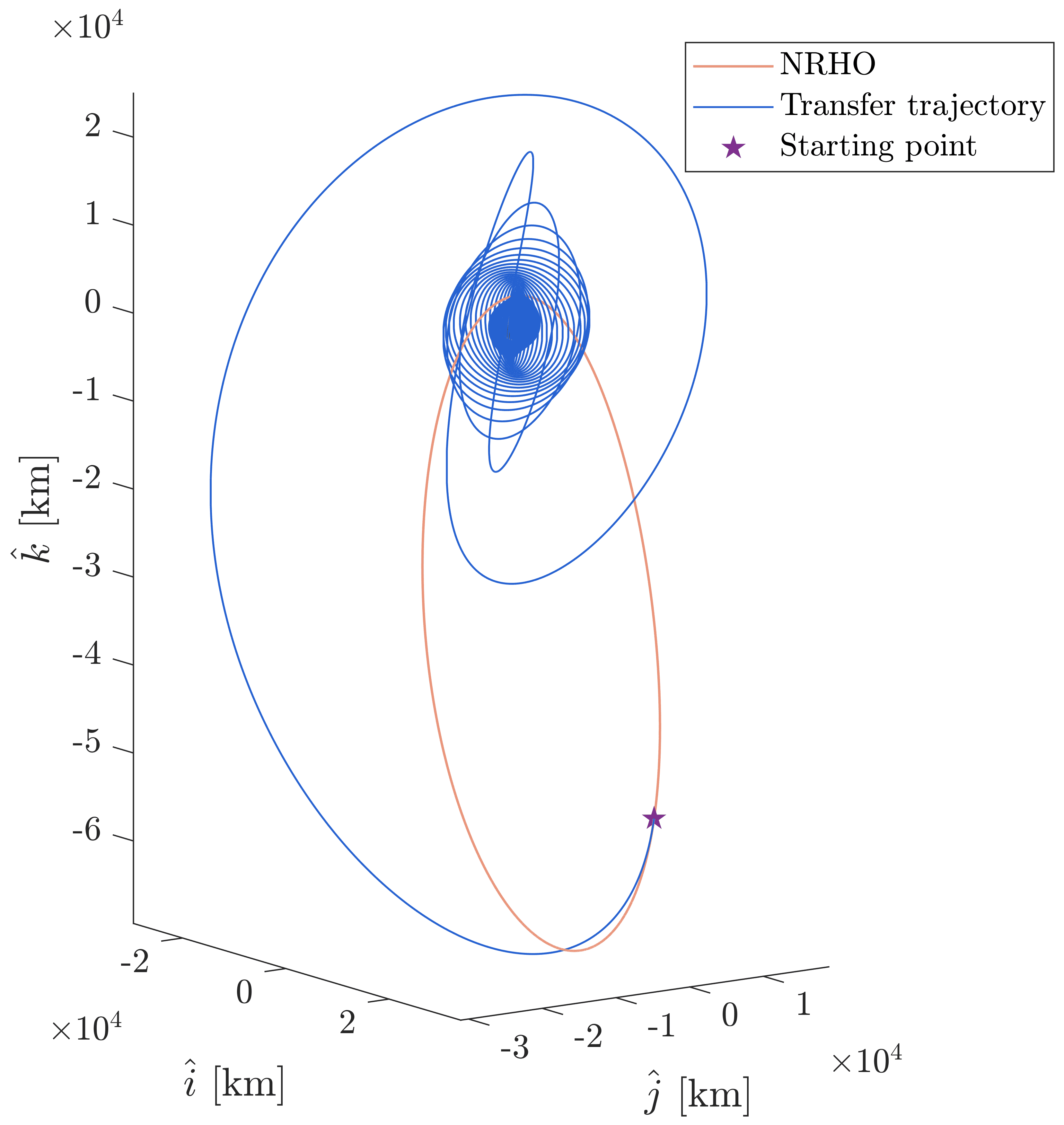}
    \caption{Transfer trajectory via feedback control in Moon-centered
synodic frame}
    \label{trajNLOC}
\end{figure}

\begin{figure}[H]
     \includegraphics[width=0.95\columnwidth]{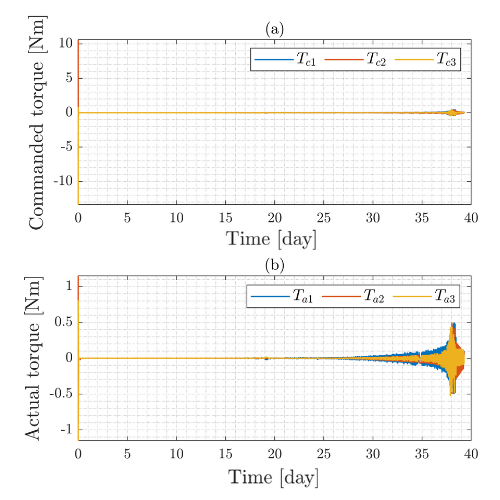}
    \caption{Time evolution of (a) the commanded torque and (b) the actual torque}
    \label{torquesNominal}
\end{figure}

\begin{figure}[H]
     \includegraphics[width=0.93\columnwidth]{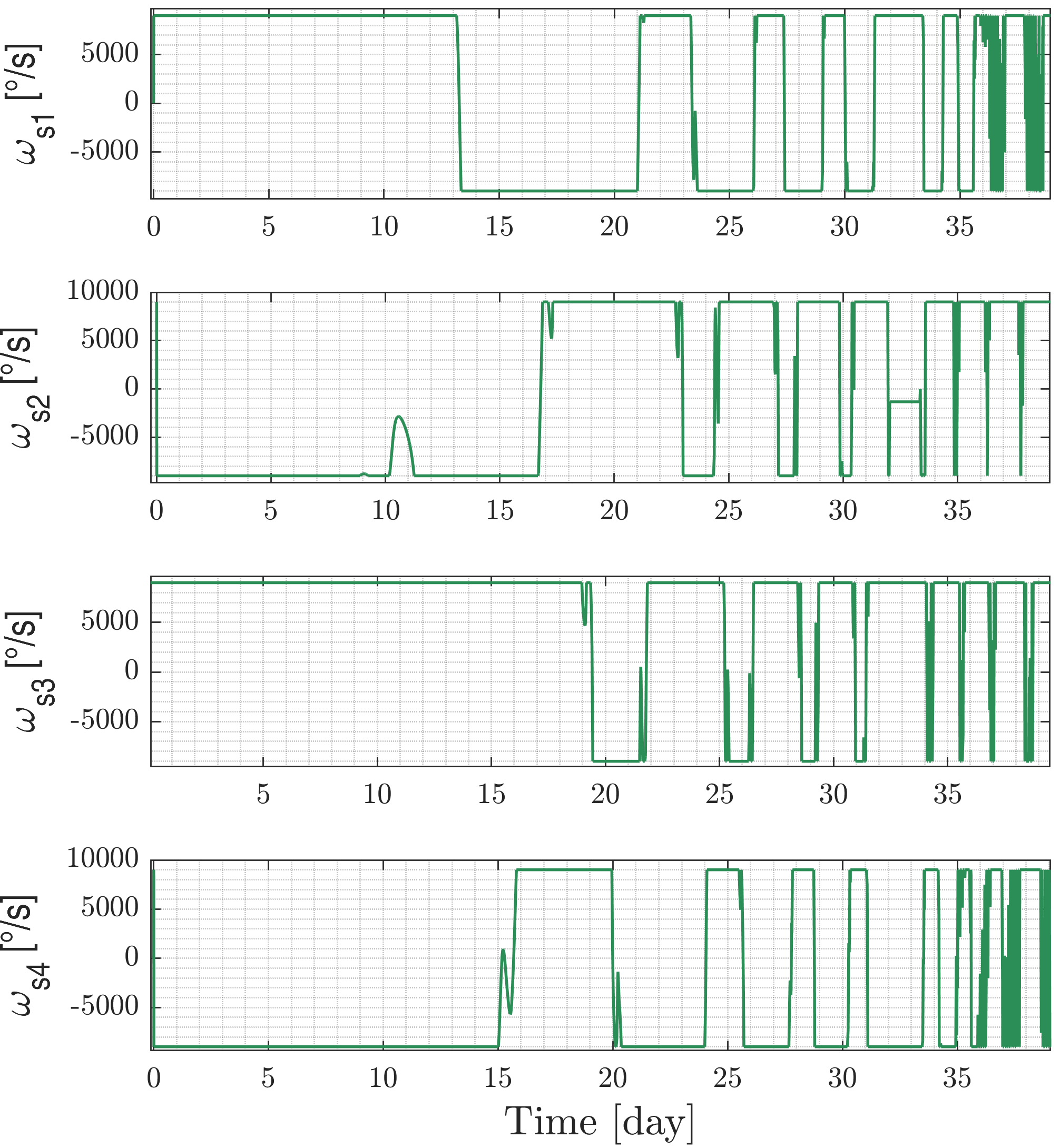}
    \caption{Time evolution of the wheels’ angular rates}
    \label{wheelsnominal}
\end{figure}

The space vehicle is modeled as a rigid body, specifically with an inertia matrix identical to that of the ATV \cite{amadieu1999automated}. The propulsion parameters are the same as in Eq. (\ref{proppar}) and the selected initial epoch for the transfer trajectory is the initial time $t_0$ associated with the optimal solution.\\
\indent As specified in Section \ref{sectionwheel}, a pyramidal array of reaction wheels is chosen for the attitude control system. The wheel's maximum angular speed and acceleration must be considered \cite{RW},
\begin{equation}\label{satu1}
\begin{split}
    \omega_{si}^{max}  &= 9000 \, \, \cfrac{\rm degree}{\rm s}\\
    \dot{\omega}_{si}^{max} &= 350 \, \,  \, \, \cfrac{\rm degree}{\rm s^2}.
\end{split}
\end{equation}
\indent As a result of systematic research aimed at finding the optimal gains in terms of flight time, the following values (appearing as diagonal terms of $\bm K$) are selected: $k_1 = 0.0338$, $k_2 = 813.373$ and $k_3 = 1.286$. For the attitude control law the two diagonal matrices $\bm A$ and $\bm B$ have the following terms: $a_1 = a_2 = a_3 = 5000 \, \, {\rm{s^2}} = a$, $b_1 = b_2 = b_3 = 100 \, \, {\rm {s}} = b$. The resulting time of flight equals $ \Delta t = 38 \, \, {\rm d} \, \, 20 \, \, {\rm hrs} \, \, 3 \, \, {\rm min} \, \, 10 \, \, {\rm s} $ (2 days and 15 hours more than the optimal solution). The spacecraft enters the desired orbit with a mass ratio of 0.9475 (propellant consumption equal to 5.23\% of the initial mass, slightly more than the percentage of the minimum-time trajectory, which is 5.12\%).\\
\indent The accuracies for the feedback orbital transfer problem are selected to finely satisfy the constraints. This selection determines the time of flight associated with feedback control, which is then compared to the value obtained from optimal control. These accuracies for the boundary conditions are
\begin{equation}\label{accuraciesNLOC}
    \begin{split}
    {\Psi}_1 &\equiv \left|x_1-p_d\right| \leq 10^{-7} \, \, \, \, \rm DU\\
    {\Psi}_2 &\equiv \left |x_2^2+x_3^2-e_d^2 \right | \leq 10^{-6} \\
    {\Psi}_3 &\equiv \left |x_4^2+x_5^2-\tan^2{\cfrac{i_d}{2}}\right | \leq 10^{-8}.
\end{split}
\end{equation}
\indent Figure \ref{trajNLOC} shows the transfer trajectory. Its behavior resembles the optimal trajectory in Fig. \ref{trajopt}. Likewise, the time variations of the orbit elements (not reported here for the sake of brevity) are similar to those obtained in the optimal solution. Figure \ref{torquesNominal} shows the time evolution of the actual torque applied to the system, which is about ten times lower than the commanded one. The difference between the two torques is due to the maximum torque deliverable by the wheel array. Figure \ref{wheelsnominal} shows the time history of the wheels’ angular rates, and one can notice that their magnitude is saturated according to Eq. (\ref{satu1}).
% \newpage
% \begin{figure}[h]
%      \includegraphics[width=\columnwidth]{torquesNominal (2).png}
%     \caption{Time evolution of (a) the commanded torque and (b) the actual torque}
%     \label{torquesNominal}
% \end{figure}
% \begin{figure}[h]
%      \includegraphics[width=\columnwidth]{omegaSthenew.png}
%     \caption{Time evolution of the wheels’ angular rates}
%     \label{wheelsnominal}
% \end{figure}

\subsection{Numerical results in nonnominal flight conditions}
Feedback control is advantageous as a real-time guidance strategy, particularly in nonnominal flight conditions. In this study, a stochastic temporary failure is introduced to test the guidance and control architecture. Specifically, a failure in the propulsion system is modeled, with randomly chosen initial time and duration. 


\begin{figure}[H]
\includegraphics[width=\columnwidth]{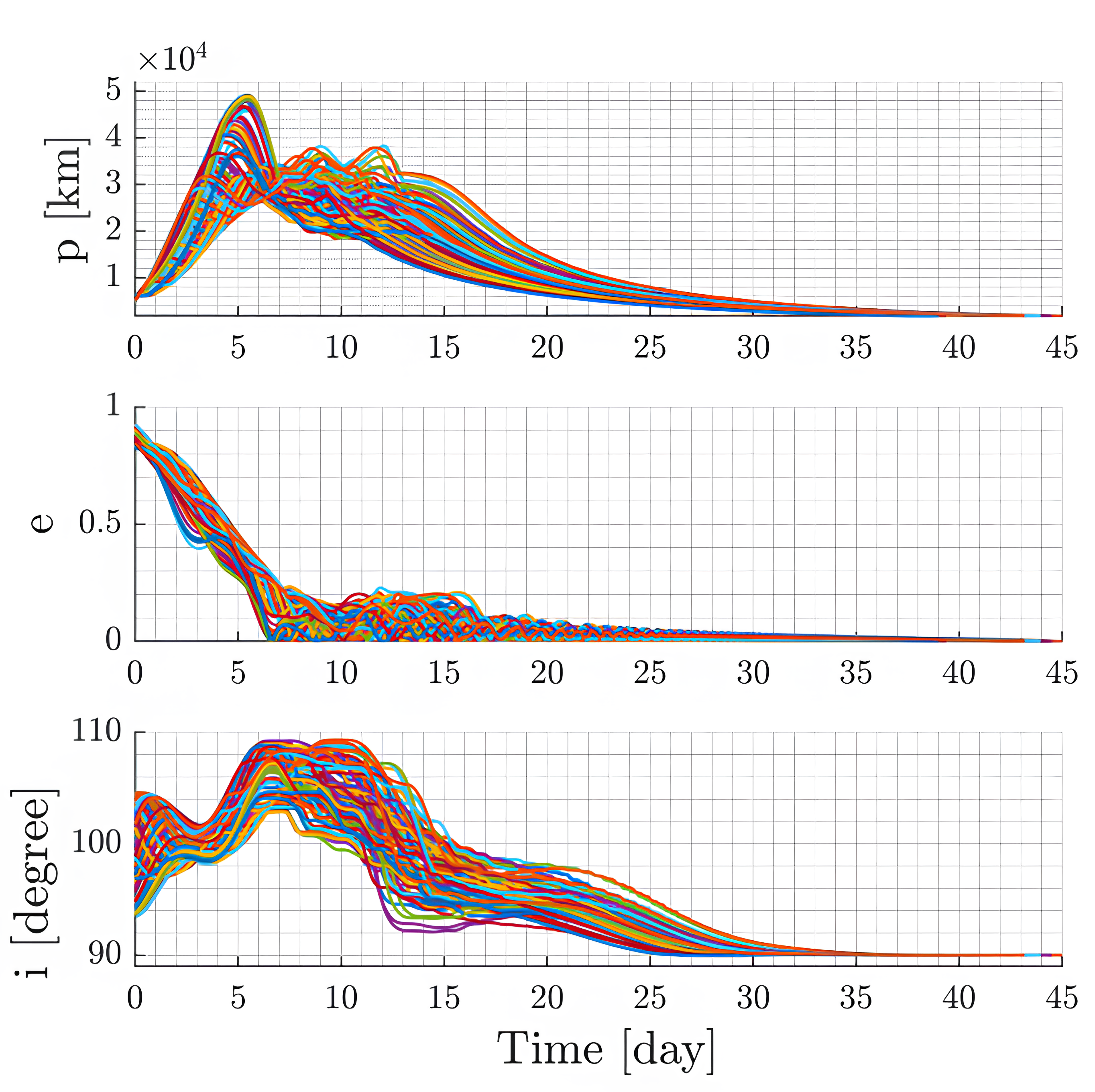}
   \caption{Controlled orbit elements, 100 Monte Carlo simulations}
    \label{MONTE}
\end{figure}

\begin{figure}[H]
   \includegraphics[width=1\columnwidth]{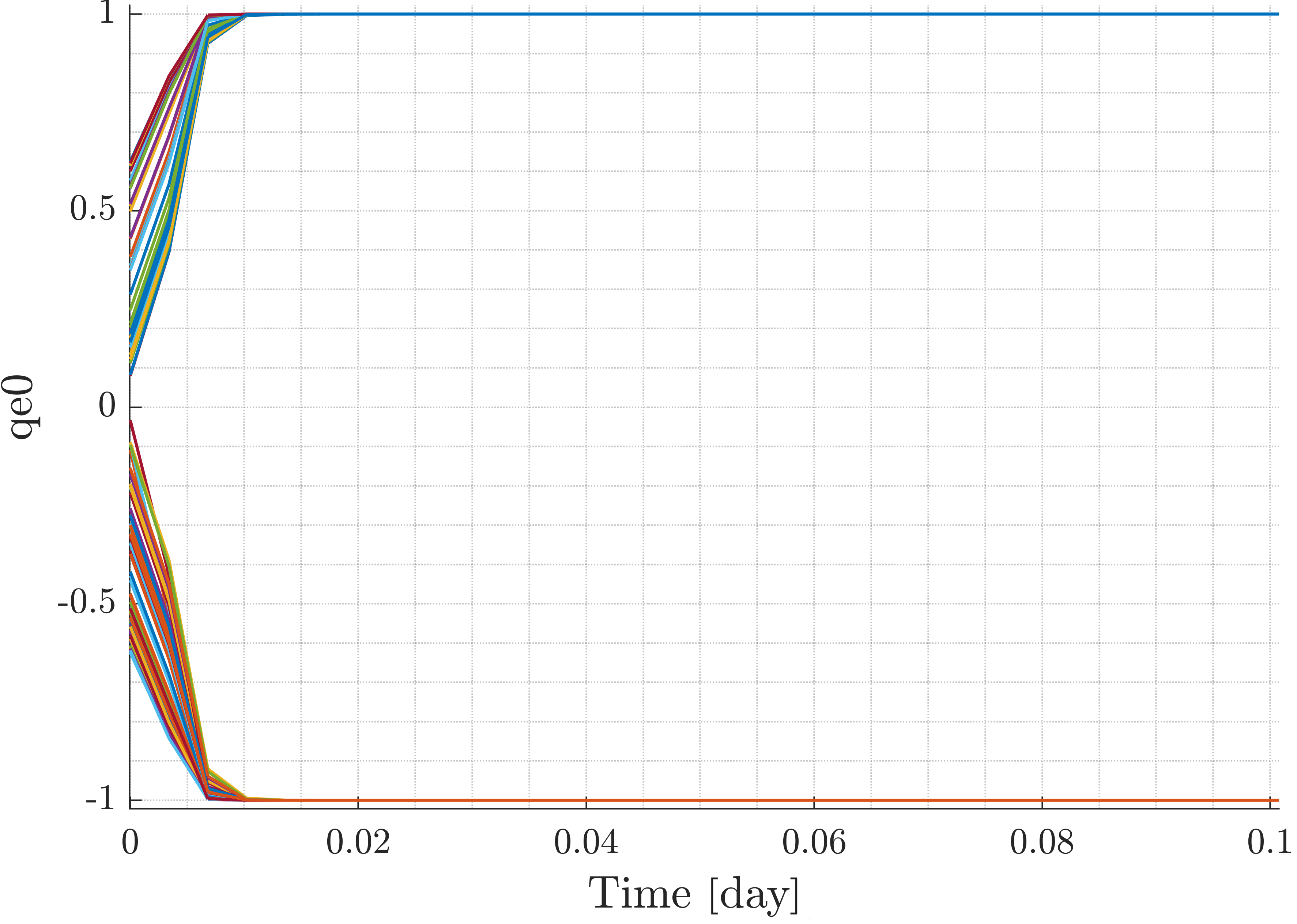}
   \caption{Scalar part of the error quaternion, 100 Monte Carlo simulations (zoom on the first 2 hours of flight)}
    \label{MONTEqe0}
\end{figure}

The duration of the failure ranges uniformly in the interval [1, 5] days, and the time of occurrence follows a uniform distribution in the interval [7, 10] days. Additionally, the starting point of the transfer trajectory is randomly selected within one period of Gateway. \noindent The initial attitude is chosen randomly, by considering an orientation associated with stochastic eigenaxis and eigenangle. The former is uniformly distributed on a unitary sphere, whereas the latter has uniform distribution in the interval [0, 180] degree. The initial angular velocity vector is set to 0.\\
\indent A Monte Carlo campaign composed of 100 simulations was run. Figures \ref{MONTE} and \ref{MONTEqe0} show the related numerical results, which point out the effectiveness of the comprehensive architecture in driving the spacecraft toward the desired orbit despite the failure. All the relevant results are summarized in Table \ref{tabmonte}. In Figure \ref{MONTEqe0} the time evolution of the scalar part of the error quaternion is portrayed; it is clear that after a few hours of flight, the quaternion matches the commanded one (i.e., $q_{e0} = \pm 1$).\\

\begin{table}[H]
\centering
\begin{tblr}{rows={rowsep=0.2ex}, columns={m}}
\hline
Parameter & Mean Value & Std Deviation \\ 
\hline
Time of flight [days] & $42.098$ &  $1.674$\\
Semilatus rectum [km] & $1837.399$ &  $1.743\cdot 10^{-4}$\\
Eccentricity & $6.710 \cdot 10^{-4}$ & $2.912 \cdot 10^{-4}$ \\
Inclination [degree] & $90.000$ & $6.922 \cdot 10^{-10}$ \\
Final mass ratio & $9.467 \cdot 10^{-1}$ & $1.429 \cdot 10^{-3}$\\ \hline
\end{tblr}
\caption{Results from Monte Carlo campaign}
\label{tabmonte}
\end{table}

\begin{figure}[H]
\includegraphics[width=\columnwidth]{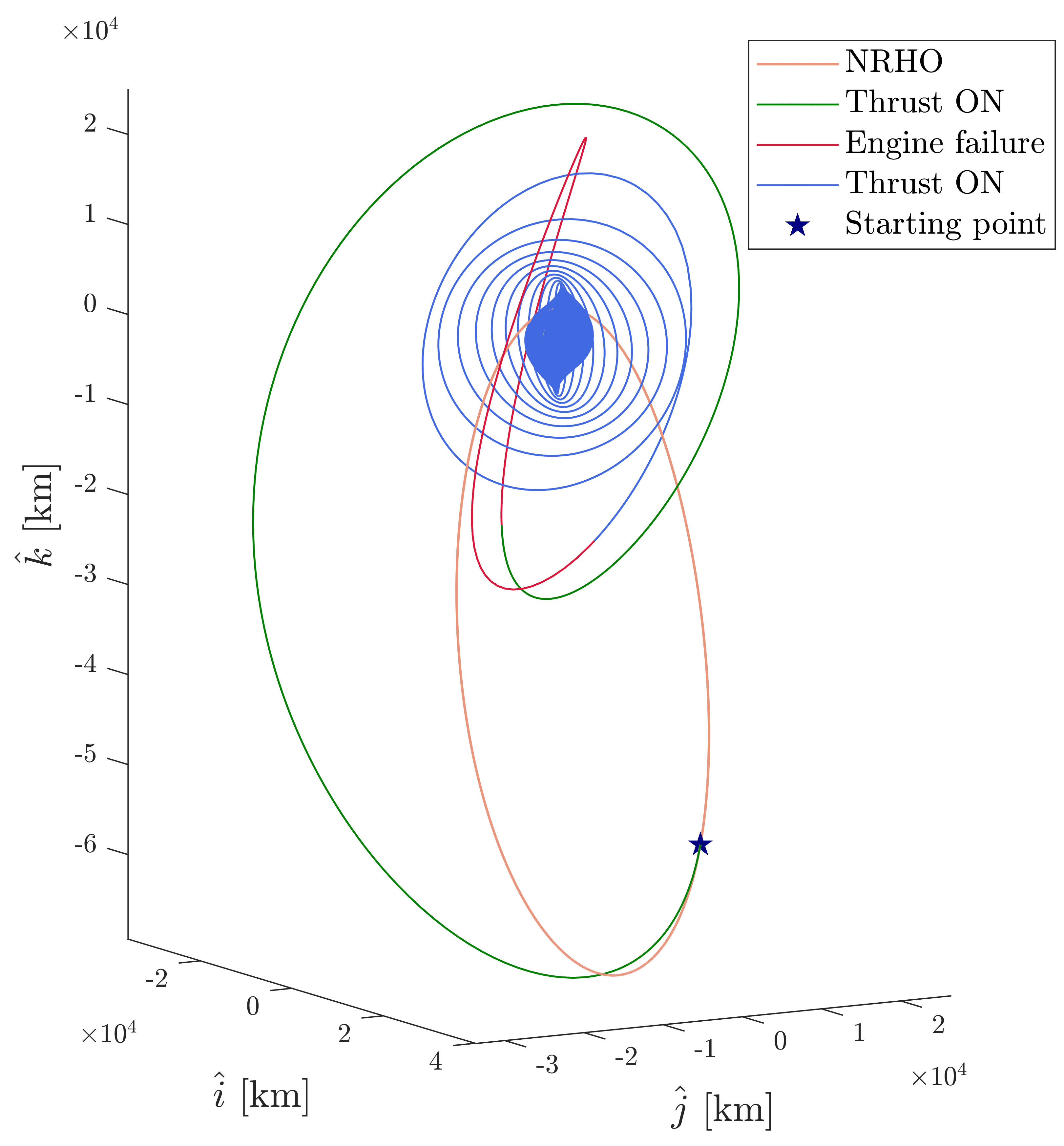}
   \caption{Transfer trajectory with 9 flight days before failure and 4 days of engine
failure}
    \label{trajFAILNLOCt0opt}
\end{figure}

\begin{figure}[H]
\includegraphics[width=0.93\columnwidth]{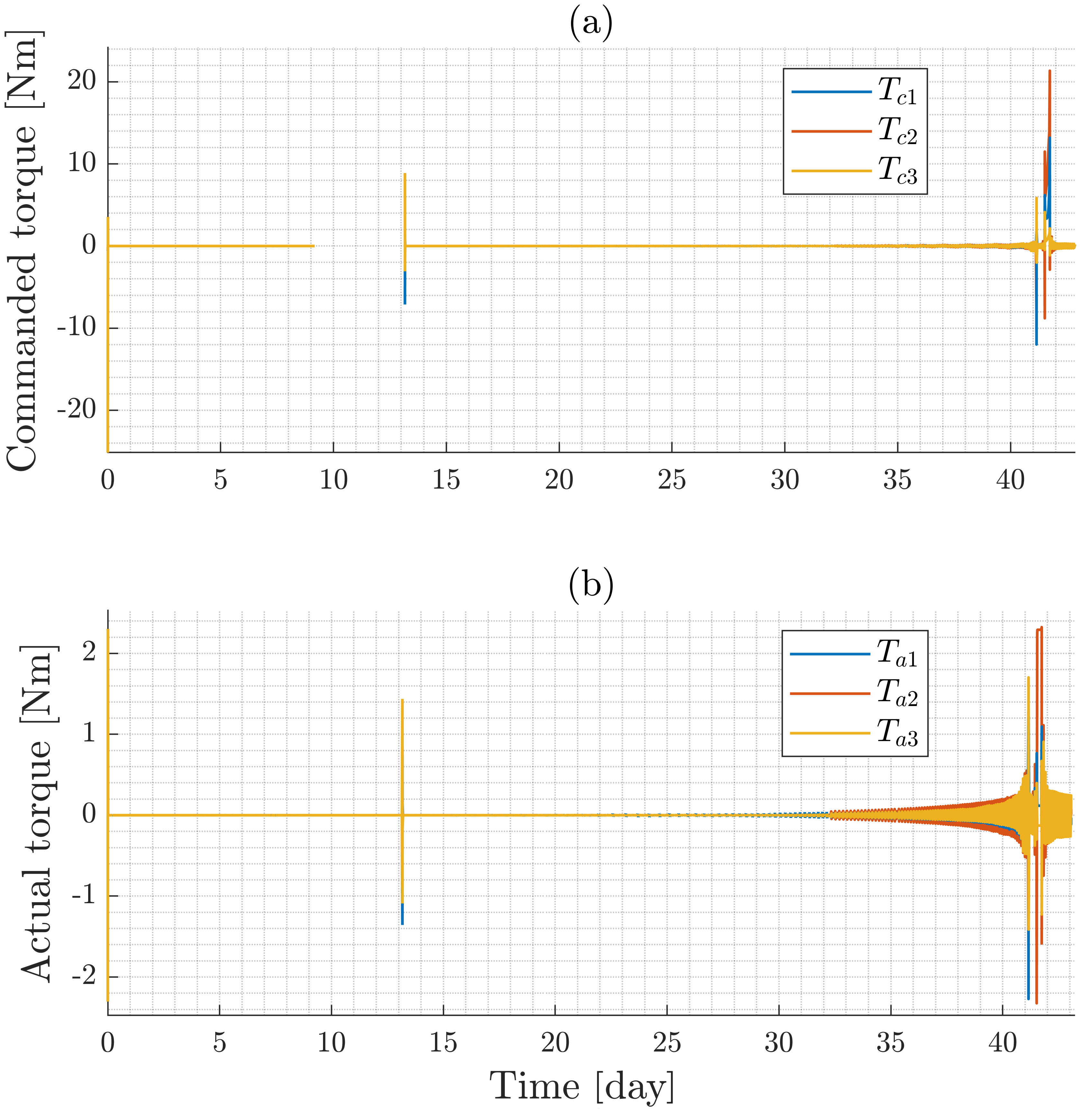}
   \caption{Time evolution of (a) the commanded (b) the actual torques}
    \label{TORQUE}
\end{figure}

\begin{figure}[H]
\centering
   \includegraphics[width=0.93\columnwidth]{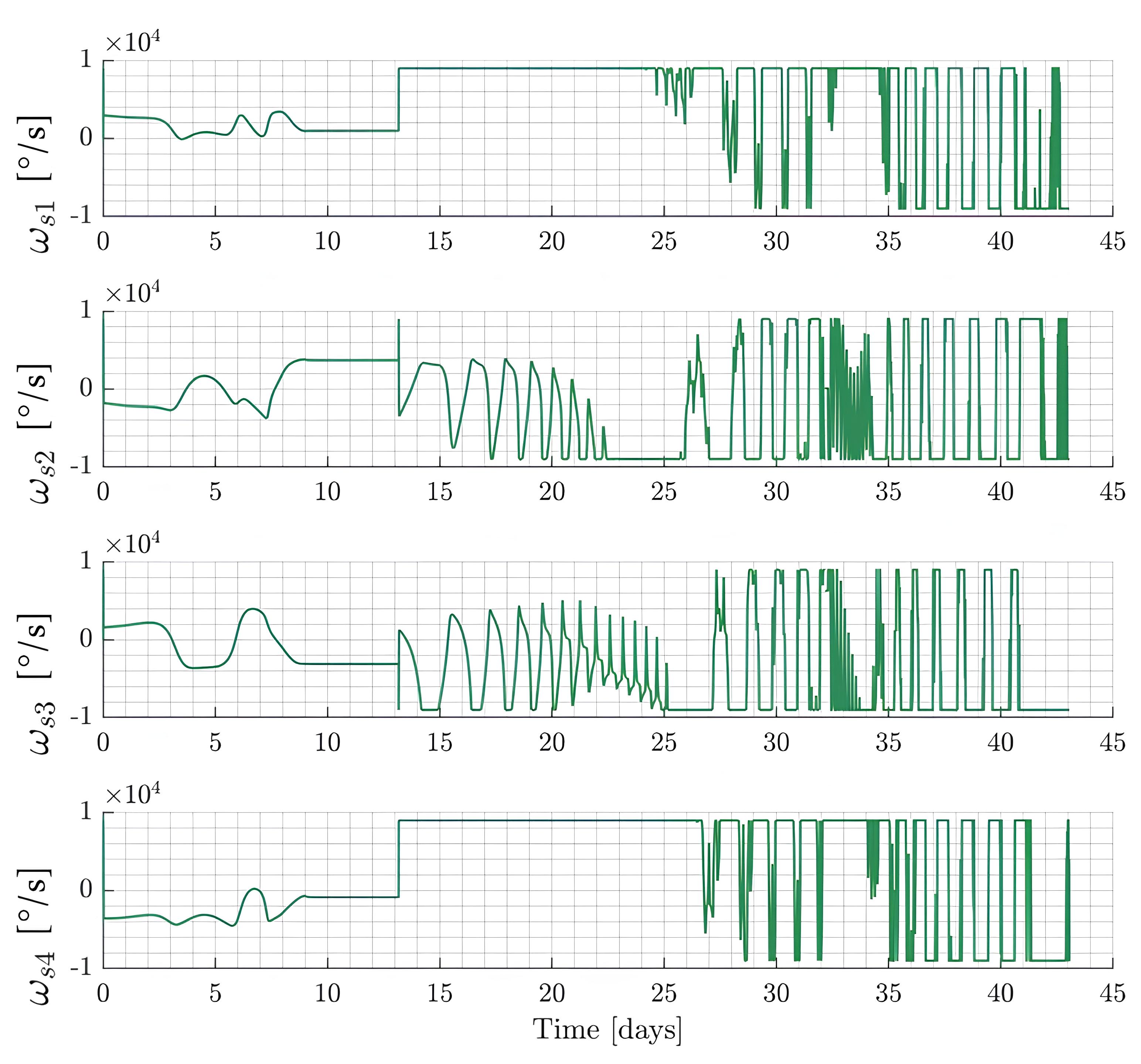}
   \caption{Time evolution of the wheels’ angular rates}
    \label{OMEGAS}
\end{figure}

\indent Figures \ref{trajFAILNLOCt0opt} through \ref{OMEGAS} provide a detailed glimpse into one of the 100 Monte Carlo simulations performed for this study. The specific nonnominal scenario under consideration involves 9 flight days before the propulsion system failure, followed by 4 days of engine inactivity. 
Figure \ref{trajFAILNLOCt0opt} displays the trajectory traveled by the spacecraft, including both thrust and coast arcs. 

In Figure \ref{TORQUE}, both the commanded and actual torque are depicted. During the failure days, the torque is zero since no commanded thrust direction is computed. \indent Furthermore, Figure \ref{OMEGAS} illustrates the time history of the angular rates of the four wheels. During the engine failure period, the wheels maintain a constant speed.

% \newpage

\indent Analyzing the results from this specific simulation is essential to evaluate how the guidance and control architecture performs under nonnominal conditions. The randomness of the initial time, duration, and starting point of the transfer trajectory allows accessing a wide range of possible scenarios and testifies to the effectiveness of the feedback control approach. These simulations provide valuable insights into the performance and reliability of the guidance and control architecture, proving its real-time applicability in nonnominal flight conditions. \\
\indent It is worth remarking that navigation errors are not modeled, because this is beyond the scope of this study, and the state of the spacecraft is assumed to be perfectly known. In a realistic scenario, navigation in cislunar space requires accurate real-time state information. Relying solely on Earth-based observations may not be sufficient, particularly over the lunar far side \cite{hesar2015lunar}. Thus, an autonomous system becomes necessary to safely and efficiently navigate a spacecraft in the vicinity of the Moon, especially for long-duration missions such as the transfers studied in this work. Several researches present effective strategies for spacecraft navigation in cislunar space \cite{christian2015optical,franzese2019autonomous,zhang2021guidance}. In particular, the sensor data fusion approach proposed by Zhang at al. \cite{zhang2021guidance}, which integrates data from star trackers, sun sensors, inertial measurement units, radars, laser altimeters, and velocimeters, might be applicable for lunar transfers. Moreover, the availability of a communication and navigation constellation around the Moon would greatly facilitate real-time operations and precise onboard positioning \cite{zanotti2024high}. With this regard, NASA, ESA, and JAXA have unveiled their project on lunar satellite constellation systems \cite{esper2022draft,giordano2021moonlight,murata2022lunar} in support to cislunar missions and the Artemis program.

\section{{Concluding remarks}}
This research identifies the minimum-time low-thrust transfer between Gateway's Near-Rectilinear Halo Orbit and a Low Lunar Orbit. Optimal control theory offers a trajectory design framework capable of solving the challenging orbit transfer problem at hand. For the numerical solution, this research employs the indirect heuristic technique, based on the joint use of all the necessary conditions for optimality and a heuristic algorithm (i.e., Particle Swarm Optimizer).\\
\indent Moreover, this study designs and numerically tests an overall guidance, control, and actuation architecture, in the context of Gateway transfer problem. Nonlinear feedback algorithms are implemented and employed for both orbit control and attitude tracking, while attitude actuation is demanded to a pyramidal array of reaction wheels. Effectiveness of the complete architecture is shown and is related to accurate tracking of the desired attitude. This is found in relation to the thrust direction, identified through a feedback, explicit-type guidance law based on nonlinear orbit control, with no need for a pre-computed offline trajectory. A slight increase in flight time (with respect to the optimal solution) is the price to pay for a 6-degree-of-freedom dynamical model of the spacecraft, and the implementation of feedback guidance and attitude control. A Monte Carlo campaign further confirms the effectiveness and precision of the architecture, also in the presence of temporary engine failure and random departure point from Gateway. 

\section*{Appendix A: Approximate Analytical Solution}\label{appendixA}
%\subsection*{Approximate analytical solution} 

In this appendix, an analytical approach is introduced and utilized to derive an estimate of the time of flight for the low-thrust orbit transfer. This approach relies on two fundamental simplifying approximations:
\begin{itemize}
    \item The initial and final orbits are circular, with final orbit radius greater than the initial one ($R_f > R_0$)
    \item Thrust vector parallel to the velocity vector at all times ($\underrightarrow{T} \parallel \underrightarrow{v}$)
\end{itemize}
\noindent In this case, the rate of change of the specific energy $\epsilon$ is
\begin{equation}
    \dot{\epsilon} = \frac{\underrightarrow{\bm T}}{m} \cdot \underrightarrow{\bm v} = a_T \, v \, \, \, \, \, \, \, \, \, \, \, \,\text{where $a_T = \frac{T}{m}$}.
\end{equation}

\noindent The specific energy is described by the following relation:

\begin{equation}\label{energyeq}
    \epsilon = -\frac{\mu}{2a}
\end{equation}

\noindent with $\mu$ the gravitational parameter of the attractive body, and $a$ the semimajor axis of the transfer trajectory. Differentiation with respect to time of Eq. \ref{energyeq} yield

\begin{equation}\label{energydot}
    \dot{\epsilon} = \frac{\mu}{2 \, a^2} \, \dot{a} = a_T \, v .
\end{equation}

\noindent Moreover, because the eccentricity is modest, one can use the following approximate relation:

\begin{equation}\label{veloxapprox}
    v \simeq \sqrt{\frac{\mu}{a}}.
\end{equation}

\noindent Substitution of (\ref{veloxapprox}) into (\ref{energydot}) yields 

\begin{equation}\label{tobeinte}
    \dot{a}=\frac{2 \, a_T}{\sqrt{\mu}}\,a^{\frac{3}{2}}.
\end{equation}

\noindent By integrating both sides of Eq. (\ref{tobeinte}) one gets

\begin{equation}
    \int_{a_0}^{a_f}a^{-\frac{3}{2}} {\rm d} a = \int_{t_0}^{t_f}\frac{2 \, a_T}{\sqrt{\mu}}{\rm d} t.
\end{equation}

\noindent Taking into account the linear time evolution of the mass ratio in Eq. (\ref{x7anal}), and considering the magnitude of the thrust acceleration, as per the PMP
\begin{equation}
    a_T = \frac{u_T^{(max)}}{x_7}
\end{equation}

\noindent one gets the approximate analytical solution of the time-of-flight

\begin{equation}
    \Delta t = \frac{c}{u_T^{(max)}}\left(1-e^{-\cfrac{\sqrt{\mu}}{c}\left(\cfrac{1}{\sqrt{a_0}}-\cfrac{1}{\sqrt{a_f}}\right)}\right) 
\end{equation}
\noindent where
\begin{equation}
\begin{cases}
        R_0 = a_0\\
        R_f = a_f.
    \end{cases}
\end{equation}

With the aim of adapting the approximate solution to the case of interest, a fictitious circular orbit with semimajor axis (i.e. radius) equal to the mean semimajor axis of the NRHO is used. The resulting approximate analytical time of flight is 

\begin{equation}\label{tfexpected}
    \Delta t = 30 \, \, \text{d} \, \, 15 \, \, \text{h} \, \, 20 \, \, \text{min} \, \, 10 \, \, \text{s}. 
\end{equation}

\noindent Although the presented analytical method only considers a single attractive body (i.e. neglecting the third body perturbation), it is rather reasonable to expect that the transfer time is underestimated by this approach. This is due to the fact that the three-dimensionality of the orbit transfer is neglected (i.e. the orbital plane changes required for the transfer path to reach the desired orbit are not taken into account).

\section*{Acknowledgment}
C. Pozzi, A. Beolchi, and E. Fantino acknowledge financial support from projects CIRA-2021-65/8474000413
(Khalifa University of Science and Technology’s internal
grant), 8434000368 (6U Cubesat mission funded by Khalifa University of Science and Technology and Al Yah Satellite
Communications Company - Yahsat) and ELLIPSE/8434000533 (Technology Innovation Institute).

\bibliographystyle{ieeetr}
\bibliography{paper.bib}

\end{document}